\documentclass{aa}  
\usepackage{supertabular}
\usepackage{siunitx}
\usepackage{graphicx}
\usepackage{txfonts}
\usepackage{float}
\usepackage{lscape}
\usepackage{supertabular}  
\usepackage{longtable}
\usepackage{hyperref}
\usepackage{float}
\usepackage{tabularx}
\usepackage{adjustbox}
\usepackage{xcolor}
\hypersetup{
    colorlinks=false,
    linkcolor=blue,
    filecolor=magenta,      
    urlcolor=cyan,
    pdftitle={Overleaf Example},
    pdfpagemode=FullScreen,
    }
\begin{document} 

   \title{An overview of desorption parameters of \\Volatile and Complex Organic Molecules}
    \subtitle{A systematic dig on experimental literature}
   
\titlerunning{An overview of thermal desorption parameters of Volatile and Complex Organic Molecules}

   \author{N.F.W. Ligterink
          \inst{1} \&
          M. Minissale
          \inst{2}
          }

   \institute{Space Research \& Planetary Sciences, University of Bern, Switzerland\\
              \email{niels.ligterink@unibe.ch}
              \and
              Aix Marseille Univ, CNRS, PIIM, Marseille, France\\
              \email{marco.minissale@univ-amu.fr}
             }

   \date{Received March 16, 2023; accepted June 15, 2023}
 
  \abstract
   {Many molecules observed in the interstellar medium are thought to result from thermal desorption of ices. Parameters such as desorption energy and pre-exponential frequency factor are essential to describe the desorption of molecules. Experimental determinations of these parameters are missing for many molecules, including those found in the interstellar medium.}
   {The objective of this work is to expand the number of molecules for which desorption parameters are available, by collecting and re-analysing experimental temperature programmed desorption data that are present in the literature.}
   {Transition State Theory (TST) is used in combination with the Redhead equation to determine desorption parameters. Experimental data and molecular constants (e.g., mass, moment of inertia) are collected and given as input.}
   {Using the Redhead-TST method, the desorption parameters for 133 molecules have been determined. The Redhead-TST method is found to provide reliable results that agree well with desorption parameters determined with more rigorous experimental methods. The importance of using accurately determined pre-exponential frequency factors to simulate desorption profiles is emphasised. The large amount of data allows to look for trends, the most important is the relationship log$_{10}$($\nu$) = 2.65ln($m$) + 8.07, where $\nu$ is the pre-exponential frequency factor and $m$ the mass of the molecule. }
   {The data collected in this work allow to model the thermal desorption of molecules and help understand changes in chemical and elemental composition of interstellar environments.}


   \maketitle
%

\section{Introduction}

Desorption of molecules from and adsorption of gaseous species on a surface play a pivotal role in regulating physical processes and setting the chemical composition of environments in the interstellar medium (ISM), star- and planet-forming regions, and solar system objects. For example, the chemical composition of hot core and corinos, compact regions of warm and molecule-rich gas surrounding protostars, are largely explained by the desorption of species from ice-coated dust grains \citep[e.g.,][]{ligterink2018b,ligterink2020b,ligterink2021,ligterink2022,bogelund2019,gorai2020,yang2021,hsu2022,nazari2022b,bianchi2022,zhang2023}. The temperature of interstellar grains and their ice mantles dictates which molecules adsorb and consequently take part in chemical reactions \citep{jin2020,garrod2022}. The molecular composition of comets like 67P/Churyumov-Gerasimenko is largely set by which species have remained frozen since its formation or have frozen-out on its surface since \citep[e.g.,][]{mumma2011,goesmann2015,altwegg2016,rubin2019}. Frozen molecules are found on the surfaces of planets and moons in the solar system, where seasonal changes alter sublimation rates and in turn affect atmospheric processes and chemical composition \citep{fray2009}, for example on Triton \citep{bertrand2022} or Pluto \citep{johnson2021}. To interpret observational data and model physical and chemical processes, empirical equations are employed to describe the desorption/adsorption process. For these equations, molecule-specific parameters need to be known. 

The theoretical framework and experimental methods for thermal desorption studies are well described in astrophysical and astrochemical literature \citep[see reviews by][]{burke2010,minissale2022}. In short, to simulate the desorption rate the Polanyi-Wigner equation is generally used:

\begin{equation}
-\frac{d N}{d t}  = \nu_{\rm n} \cdot N ^{\rm n} \cdot {\rm exp} \left( \frac{E_{\rm des}}{T} \right),
\label{eq:polwig}
\end{equation}

where $\nu_{\rm n}$ the pre-exponential frequency factor with value molecules$^{\rm 1-n}$ s$^{-1}$ (also often denoted as $A_{\rm n}$ and where $n$ is the desorption order with n = 0, 1, 2), $N$ the surface coverage in molecules cm$^{-2}$ (also often denoted as $\theta$), n (= 0, 1, 2) the order of desorption, $E_{\rm des}$ the desorption energy in K, and $T$ the temperature of the surface. The order of desorption is given as zeroth, first, or second. Zeroth order desorption is associated with multilayer desorption, while first order desorption with (sub)monolayer desorption. The desorption energy can also be given in Joule by changing the exponent to $E_{\rm des}$ $k_{\rm B}^{-1}$ ($k_{\rm B}$ = Boltzmann constant) or in Joule mol$^{-1}$ by changing the exponent to $E_{\rm des}$ $R^{-1}$ ($R$ = ideal gas constant). Second order desorption is possible, and is for example observed with processes such as recombinative desorption, where two species react to form the desorbing product. However, as this type of desorption is hardly encountered within the astrochemical literature, second order desorption is ignored in the remainder of this publication. 

There are a variety of experimental methods to determine the desorption parameters n, $\nu$, and $E_{\rm des}$, most of them based on the Temperature Programmed Desorption (TPD) technique or a variation thereof. In a typical TPD experiment, a surface held at low temperature under vacuum conditions is exposed and covered with a given adsorbate. Next, the temperature of the surface is linearly increased. At some point the adsorbate will start desorbing and continues to do so until the adsorbate is fully removed from the surface. The release of adsorbate to the gas-phase can be traced with a variety of instruments, but usually a mass spectrometry technique is employed. The measured desorption trace can be analysed to find the desorption parameters, for example with leading edge, Redhead, heating variation, inversion, or Arrhenius analysis \citep[e.g.,][]{king1975,dejong1990b,dejong1990a,tait2005b}. 

There are a number of limitation to the experimental determination of desorption parameters, specifically relating to experimental efforts and safety. Currently, the number of molecules detected in the interstellar medium is about 270\footnote{https://cdms.astro.uni-koeln.de/classic/molecules}$^{,}$\footnote{http://astrochymist.org/astrochymist\_mole.html} \citep{mcguire2022} and more are detected every year. Because experiments are time consuming, it is challenging to keep up with the number of detections and provide desorption parameters for all species. Furthermore, some molecules are difficult to work with, either because they are chemically unstable (e.g., PH$_{2}$COOH, CH$_{3}$OOH, c-C$_{3}$H$_{4}$O) or highly toxic (e.g., HCN, CH$_{3}$NCO, H$_{2}$P(O)OH). To bridge the gap between the availability of experimentally determined desorption parameters and the needs of the community, alternative approaches are needed. Computational techniques such as Machine Learning \citep{villadsen2022}, Bayesian inference \citep{heyl2022}, DFT calculations \citep{ferrero2022,piacentino2022}, or quantum mechanical methods \citep{germain2022,bovolenta2022,tinacci2022} help fill the gap. However, there also exists a rich experimental literature of TPD experiments that have been used for the identification of molecules produced in experiments that simulate chemical processes in extraterrestrial ice (for brevity named ``chemical TPDs''), but not to assess their desorption parameters. Because this type of data is in essence the same as what is used for the determination of desorption parameters, it raises the question if chemical TPD traces can be used to determine pre-exponential factors and desorption energies and in this way contribute more of these essential parameters to the literature.

In this study, chemical TPD traces are collected from laboratory literature and analysed with a combination of Transition State Theory and the Redhead method (Redhead-TST) to determine the pre-factor and desorption energies of 133 molecules. The methods are presented in Sect. \ref{sec:method} and the resulting data in Sect. \ref{sec:result}. Implications for astrophysical and astrochemical studies are discussed in Sect. \ref{sec:implications}.

\section{Methods}
\label{sec:method}

\begin{figure*}[h]
    \centering
    \includegraphics[width=0.80\textwidth]{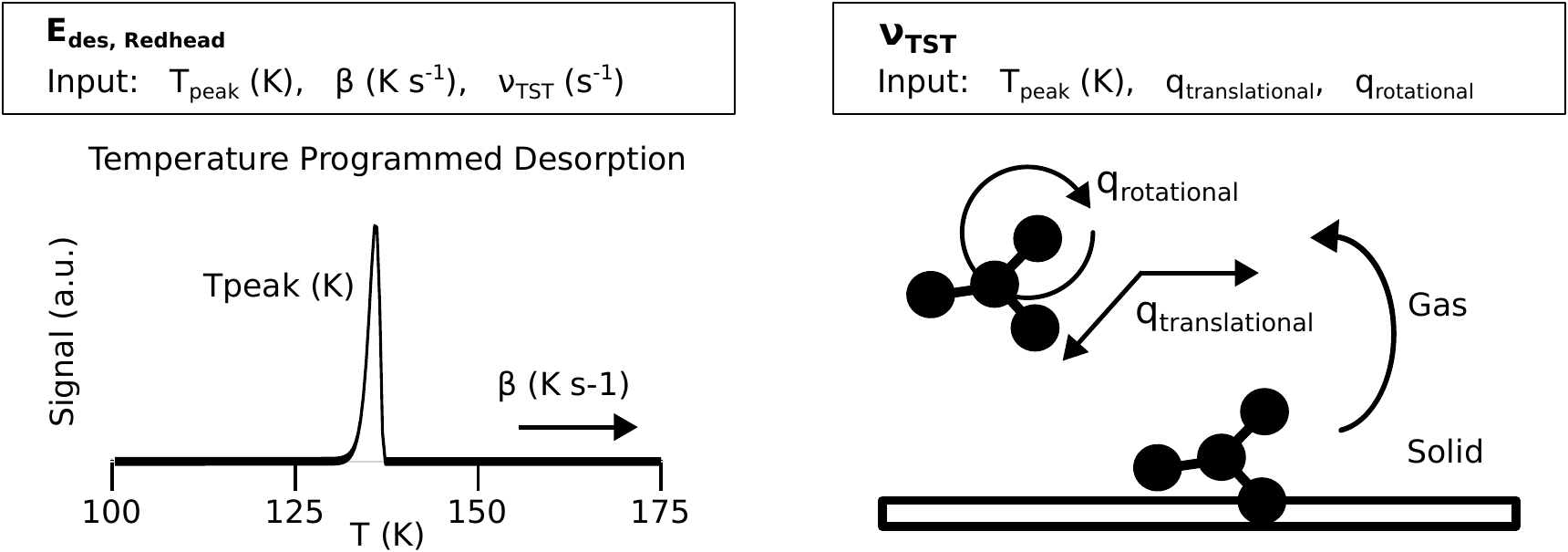}
    \caption{Visual summary of the Redhead-TST method. The pre-exponential frequency factor ($\nu$) is determined with Transition State Theory (TST). The main input are the translational ($q^{\ddagger}_{\rm tr,2D}$) and rotational ($q^{\ddagger}_{\rm rot,3D}$) partition functions, which increase as the molecule transitions from a solid to gaseous state, and $T_{\rm peak}$ of the molecule. The Redhead equation takes $T_{\rm peak}$, the heating rate $\beta$, and $\nu_{\rm TST}$ as main inputs to calculate the desorption energy $E_{\rm des}$.}
    \label{fig:summary}
\end{figure*}

This work makes use of an analysis method based on the Redhead equation and Transition State Theory, and is indicated as the Redhead-TST method throughout this manuscript. Furthermore, desorption energies and pre-exponential factors are obtained from TPD data that are used for molecule identification. The Redhead-TST method and the data set are introduced in the following sections and a visual summary is presented in Fig. \ref{fig:summary}.

\subsection{Redhead-TST formalism}
\label{sec:redhead-tst}

To determine the desorption energies of molecules from TPD data, the Redhead equation is used \citep{redhead1962,king1975}:

\begin{equation}
E_{\rm Redhead} = T_{\rm peak} \cdot \left(\ln \left( \frac{\nu_{\rm TST} T_{\rm peak}}{\beta} \right) - 3.64\right)
\end{equation}

This equation takes the peak of the desorption trace ($T_{\rm peak}$, K) in combination with the heating rate ($\beta$, K s$^{-1}$) and a pre-exponential factor ($\nu$, $s^{-1}$) to determine $E_{\rm des}$ in K energy units. This equation only applies to first order desorption, which generally applies to (sub-)monolayer coverage. It is not used for zeroth order desorption processes, which is usually the case for multilayer desorption. \citet{redhead1962} showed that for $\nu$/$\beta$ values of 10$^{8}$ -- 10$^{13}$ K$^{-1}$, the relation between $E_{\rm des}$ and $T_{\rm peak}$ is nearly linear, within a $\pm$1.5\% accuracy. For values of $\nu$/$\beta$ > 10$^{13}$ K$^{-1}$, we verified that this relationship holds by comparing literature results with those retrieved from the Redhead equation when the same parameters are used (see Sect. \ref{sec:performance}). While the Redhead equation is not the most accurate analysis method to determine desorption energies \citep{dejong1990a,dejong1990b}, the simplicity of this equation makes it very suitable for the analysis of TPD data that have been recorded for other purposes, such as the molecule identification, rather than their desorption parameters. Data of this type can be low in signal to noise ratio or have a poorly defined desorption trace shape, which make other methods, such as the leading edge analysis \citep{dejong1990a,dejong1990b} less suitable to analyse it with. 

The value for $\nu$ used in the Redhead equation is usually assumed and taken to be 10$^{12}$ -- 10$^{13}$ s$^{-1}$. While these values are suitable for small molecules and atoms, they significantly underestimate the pre-exponential factor for larger species with more degrees of freedom. In this work, the pre-exponential factor is therefore calculated by Transition State Theory (TST), following the equation:

\begin{equation}
    \nu_{\rm TST} = \frac{k_{\rm B} \cdot T_{\rm peak}}{h} \cdot q^{\ddagger}_{\rm tr,2D} \cdot q^{\ddagger}_{\rm rot,3D},
    \label{eq:tst}
\end{equation}

where $k_{\rm B}$ is the Boltzmann constant and $h$ the Planck constant. This formalism is adopted from \citet{minissale2022}, which in turn is based on work by \citet{tait2005a}. In short, TST takes the difference in rotational and translational degrees of freedom between the adsorbed and transition state into account. In equation \ref{eq:tst}, $q^{\ddagger}_{\rm tr,2D}$ and $q^{\ddagger}_{\rm rot,3D}$ are the 2D translational partition function and the 3D rotational partition function, respectively. The 2D translational partition function, because the dimension orthogonal to the surface is assumed to be common to both adsorbed and desorbed molecules. $q^{\ddagger}_{\rm tr,2D}$ is given by:

\begin{equation}
    q^{\ddagger}_{\rm tr,2D} = \frac{A}{\Lambda^2}.
\end{equation}

The parameter $A$ is the surface area of each adsorbed molecule, which is fixed to 10$^{-19}$ m$^{2}$ and is the inverse of the generally assumed number of binding sites (that is, 1$\times$10$^{15}$~cm$^{-2}$). For large molecules this value could be different, but for simplicity we adopt a uniform value. $\Lambda$ is the thermal wavelength of the molecule and calculated in the following way:

\begin{equation}
    \Lambda = \frac{h}{\sqrt{2\ \pi\ m_{\rm molecule}\ k_{\rm B}\ T_{\rm peak}}}
\end{equation}

In this equation, $m_{\rm molecule}$ is the mass of the particle in kg. Finally, the rotational partition function, $q^{\ddagger}_{\rm rot,3D}$ is given as:

\begin{equation}
    q^{\ddagger}_{\rm rot,3D} = \frac{\sqrt{\pi}}{\sigma h^3} \cdot (8 \pi^2 k_{\rm B} T_{\rm peak})^{3/2} \cdot \sqrt{I_{\rm x} I_{\rm y} I_{\rm z}}
\end{equation}

Here, $\sigma$ is the symmetry factor of the molecule and indicates the number of indistinguishable orientations of the particle. $I_{\rm x}$, $I_{\rm y}$, and $I_{\rm z}$ are the principal moments of inertia for rotation of the particle. The moments of inertia are determined using a rigid rotor approximation and chemical structures from the ChemSpider\footnote{http://www.chemspider.com} database. These structure as calculated with a Dreiding force fieled based geometry optimisation and are not a full quantum mechanical treatment. For a handful of molecules their structures were not available in this database and for these instances they have been calculated with the Avogadro\footnote{Avogadro: an open-source molecular builder and visualisation tool. Version 1.2.0 http://avogadro.cc/} software. These equations are only applicable to molecules consisting of more than two atoms. \citet{tait2005a} note that this TST method gives a good approximation of the pre-exponential factor, but can overestimate the value. Adsorbates are assumed to be immobile on the surface and therefore have no rotational or translational degrees of freedom when bound (in other words $q_{\rm ads}$ = 1). Since some molecules are found to migrate on the surface to sites with higher binding energies, they have some degrees of freedom on the surface, which results in $q_{\rm ads}$ > 1. With a mobile adsorbate, $q^{\ddagger}_{\rm tr,2D} \cdot q^{\ddagger}_{\rm rot,3D}$ / $q_{\rm ads}$ will therefore be lower, thus lowering the pre-exponential factor.

We stress that TST suffers from some limitations. One of the main approximations is that all molecules in the transition state reach the Boltzmann distribution. This could not be the case for large molecules, and it is the reasons why the TST could fail to treat large molecules. It is not easy to quantitatively define the ``large'' word but based on the results of the present work and on \citet{minissale2022}, we can tentatively claim that TST starts to fail when the molecule presents more than 20 atoms. Moreover, TST neglects quantum effects, which is important at low temperatures or when the chemical reaction involves tunnelling. We point out that this a second order limitation since, except for H$_{2}$ or D$_{2}$ or other peculiar cases, desorption occurs at temperature where classical effects overcome of some orders of magnitude quantum effects.

Several species considered in this work are salt complexes, which consist of two molecules that have engaged in an acid-base reaction and are present as a cation-anion pair. For these species, it is generally not possible to determine the moments of inertia and subsequently its pre-exponential factor with TST. To determine the desorption energies of these salts, we take their pre-factor as the sum of the pre-factors of the individual acid and base. The following values are used for individual components: 4.96$\times$10$^{15}$~s$^{-1}$ for H$_{2}$O, 1.63$\times$10$^{17}$~s$^{-1}$ for HCN, and 1.94$\times$10$^{15}$~s$^{-1}$ for NH$_{3}$ \citep{minissale2022} and 2.9$\times$10$^{17}$~s$^{-1}$ for HNCO, 3.9$\times$10$^{17}$~s$^{-1}$ for CH$_{3}$NH$_{2}$, 6.9$\times$10$^{19}$~s$^{-1}$ for NH$_{2}$COOH, 2.0$\times$10$^{18}$~s$^{-1}$ for HCOOH, 1.3$\times$10$^{19}$~s$^{-1}$ for CH$_{3}$COOH, and 1.7$\times$10$^{21}$~s$^{-1}$ for the Acetaldehyde Ammonia Trimer (AAT, this work). 

\subsection{Data set}

The data used in this paper have been collected from a wide variety of publications in the laboratory astrochemistry and surface science literature. All these publications make use of TPD experiments in combination with a mass spectrometry technique, such as quadrupole mass spectrometer (QMS) or time-of-flight MS (TOF-MS), to detect and identify molecules. 

To determine $E_{\rm Redhead}$ and $\nu_{TST}$, the peak desorption temperature $T_{\rm peak}$ and heating rate $\beta$ are collected and presented in Table \ref{tab:redhead-tst}. Table \ref{tab:mol_const} lists the relevant molecular constants mass, $\sigma$, $I_{\rm x}$, $I_{\rm y}$, and $I_{\rm z}$. The heating rate is usually indicated in the publication, but in case $\beta$ is not provided it is set to 1 K min$^{-1}$. This matches with most heating rates applied in astrochemical laboratory work, but we note that heating rates in this data set range from 0.1 K min$^{-1}$ to over 1 K s$^{-1}$. Peak desorption temperatures can textually be indicated in the publication or be determined by-eye from TPD traces presented in figures. The absolute uncertainty on the recorded temperature is generally 0.5--2.0~K, depending on the measurement technique used (e.g., thermocouple, diode). However, by-eye analysis of TPD data is inherently more inaccurate and the $T_{\rm peak}$ uncertainty is therefore uniformly set to $\pm$5~K. Combined with the uncertainties of the equations used, we apply a uniform uncertainty of $\pm$10\% on all determined desorption energies. 

Table \ref{tab:redhead-tst} also list details about the substrate material, precursor molecules (i.e., starting molecule or mixture of molecules), and the ice processing source. We note that most data are collected for molecules that are formed in-situ, instead of as pure or mixed deposited ices. This means that while compositions of the ice at the beginning of the experiment are given, by the time the TPD is started this composition has changed because new molecules have been formed. In fact, some of the listed molecules may form in reactions that are promoted by heating during the TPD. This makes characterising the binding environment of a molecule challenging, as the target molecule does not only interact with the substrate or surrounding precursor species, but potentially with a host of molecules that are formed during processing of the ice. Therefore, when the ice is processed and the target species is formed in-situ, the binding environment is assumed to consist of a combination of the substrate material (e.g., metal, carbon, etc) and a residue that is a mix of different and undefined organic molecules. All molecules are assumed to be present at (sub)monolayer coverage (1ML $\approx$ 1$\times$10$^{15}$~molecules~cm$^{-2}$), unless there is clear mention that multilayer quantities ($\gg$1ML) of product are formed, for example determined from IR spectroscopic measurements, in which case the molecule is excluded from the dataset.

It is possible that the molecule under investigation co-desorbs with another matrix species, either a precursor molecule or an new species that is abundantly formed during the processing of the ice. Because co-desorbing species have a peak desorption temperature that is governed by the matrix molecule instead of the binding of the targeted species to the surface, these molecules are excluded from the data set. For precursor molecules (e.g., H$_{2}$O, CH$_{3}$OH) desorption temperatures are often known and therefore any target molecule that has a $T_{\rm peak}$ close to this is considered to be co-desorbing and excluded. However, during processing new molecules that act as co-desorption matrix can be formed. Because the full molecular inventory is not always described, it is difficult to identify co-desorption in these cases. Only if there is mention or a strong suspicion that co-desorption is occuring the entry is omitted.

A variety of ice processing techniques are used, ranging from UV and X-ray processing to hydrogenation and electron/ion bombardment. We note that thermal processing is also a possibility, but this is only listed when it is explicitly mentioned as a step in the molecule formation process. The TPD process itself is not considered thermal processing. 

Finally, data of (sub)monolayer desorption studies presented in the literature have also been collected and are presented in Table \ref{tab:redhead-tst} as well. In some cases only $\nu_{\rm lit}$ and $E_{\rm lit}$ are presented, while in other cases sufficient information is available (i.e., $T_{\rm peak}$ and $\beta$) that $\nu_{\rm TST}$ and $E_{\rm Redhead}$ can also be determined. 

\begin{figure}[h]
    \centering
    \includegraphics[width=0.45\textwidth]{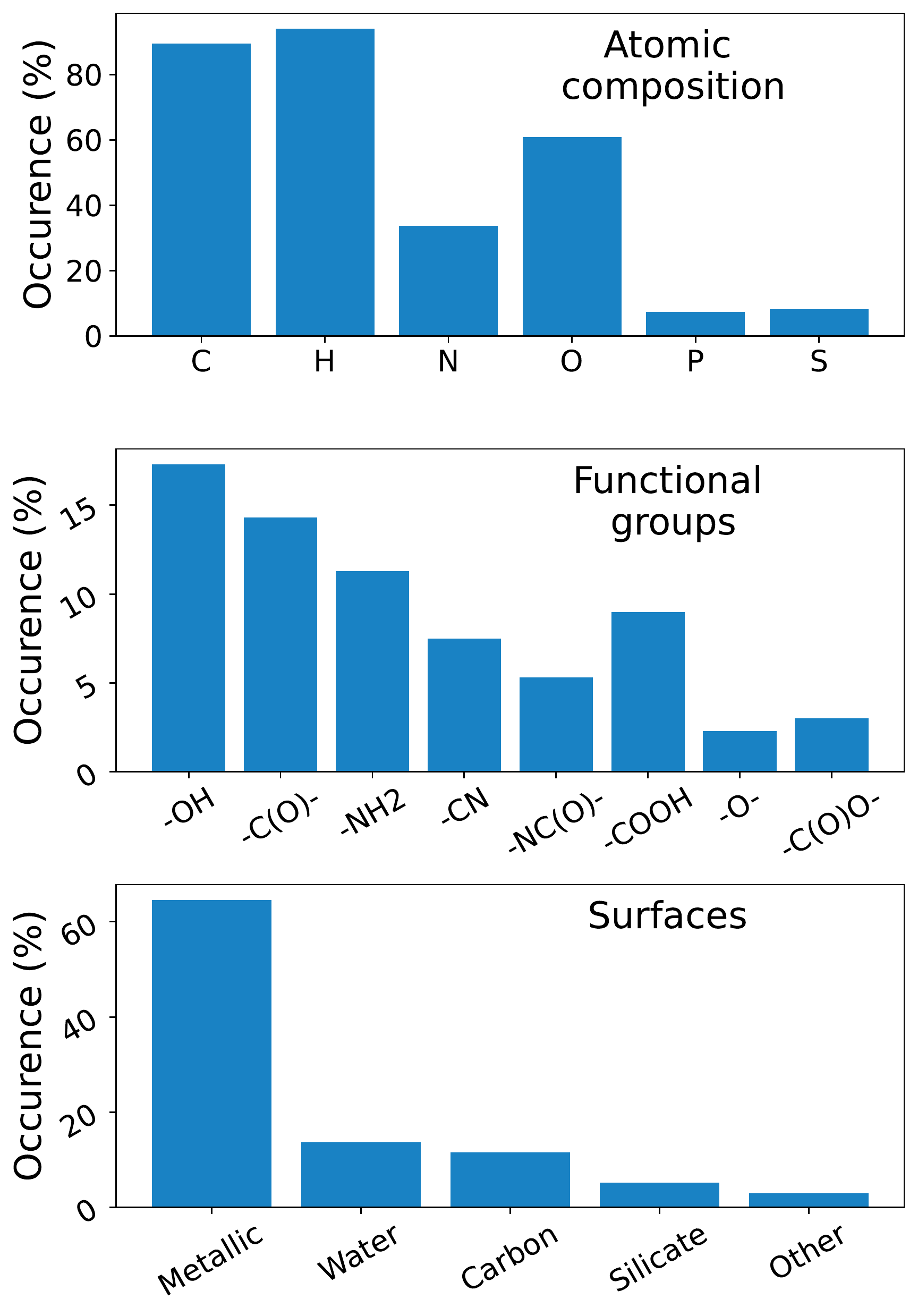}
    \caption{Occurrences of CHNOPS atoms (top panel), functional groups (middle panel), and binding surfaces (bottom panel) of the entries present in the data set used in this paper. The category ``other'' in the bottom panel includes surfaces like KBr and MgF$_{2}$ windows.}
    \label{fig:content}
\end{figure}

\section{Results and Discussion}
\label{sec:result}

In Table \ref{tab:redhead-tst} data of 133 molecules from 132 publications have been collected, for a total of 328 entries. Information on the CHNOPS elemental composition, functional groups, and binding surfaces in the data set are presented in Fig. \ref{fig:content}. The vast majority of molecules contain one or more carbon and hydrogen atom, while a substantial fraction contains at least one oxygen atom. About one third of the molecules included contains at least one nitrogen atom, while phosphorus and sulfur are present in a minor percentage of the data set. In terms of functional groups, we find that mostly alchols, aldehydes or ketones, and amines are covered in the data set. Contrary to what is found in the ISM \citep[e.g.,][]{cernicharo2020,mcguire2020,marcelino2021,lee2021a,lee2021b,loomis2021,lee2022,cernicharo2022}, cyanides only make up a small portion of this data set and ethers and formates represent the smallest fraction of the functional groups. 

This figure highlights the large availability of data experimental data on oxygen-bearing molecules, but relatively few molecules are covered that contain nitrogen, phosphorus, or sulfur atoms. There are two primary explanations for this. First, many precursor molecules with N, P, or S, such as HCN, PH$_{3}$, and H$_{2}$S, are more difficult to work with in the laboratory or toxic and therefore avoided by experimental researchers. In turn, fewer TPD data on the formation of molecules containing these kind of atoms are available. At the same time the limited amount of data available on N, P, S-bearing species may also indicate that ice chemistry is less efficient at forming such molecules or that molecules containing these atoms are refractory and do not desorb at temperature $\leq$300~K. 

Finally, the cold surfaces for these studies are dominated by metallic ones (e.g., Au, Ag, Pt). Surfaces that are more relevant to the ISM, such as crystalline or amorphous water, highly oriented pyrolytic graphite (HOPG), graphene, silica (SiO$_{2}$), and silicates (SiO$_{4}^{4-}$) make up a smaller percentage of the data set. Only few ice chemistry experiments make use of surface that are not metallic \citep[e.g.,][]{potapov2022a}. Since many molecules in the data set are obtained from such experiments, this explains the dominance of metallic surfaces in the data set. However, as mentioned earlier, these molecules are produced in-situ, likely together with a mixture of other complex organic molecules that can remain on the surface at temperatures $\geq$300~K. Consequently, the binding surface of the target molecule will be a combination of the metallic surface and an organic molecular residue. This environment be a relevant analogue to dust grains in the ISM, which are presumably coated in a layer of organic molecules once water-ice is removed from their surfaces. 

In the following sections a deeper look will be taken at the Redhead-TST method performance and collected desorption energy and pre-exponential factor data. 

\subsection{Influence of TPD data on the Redhead-TST output}

\begin{figure}
    \centering
    \includegraphics[width=0.45\textwidth]{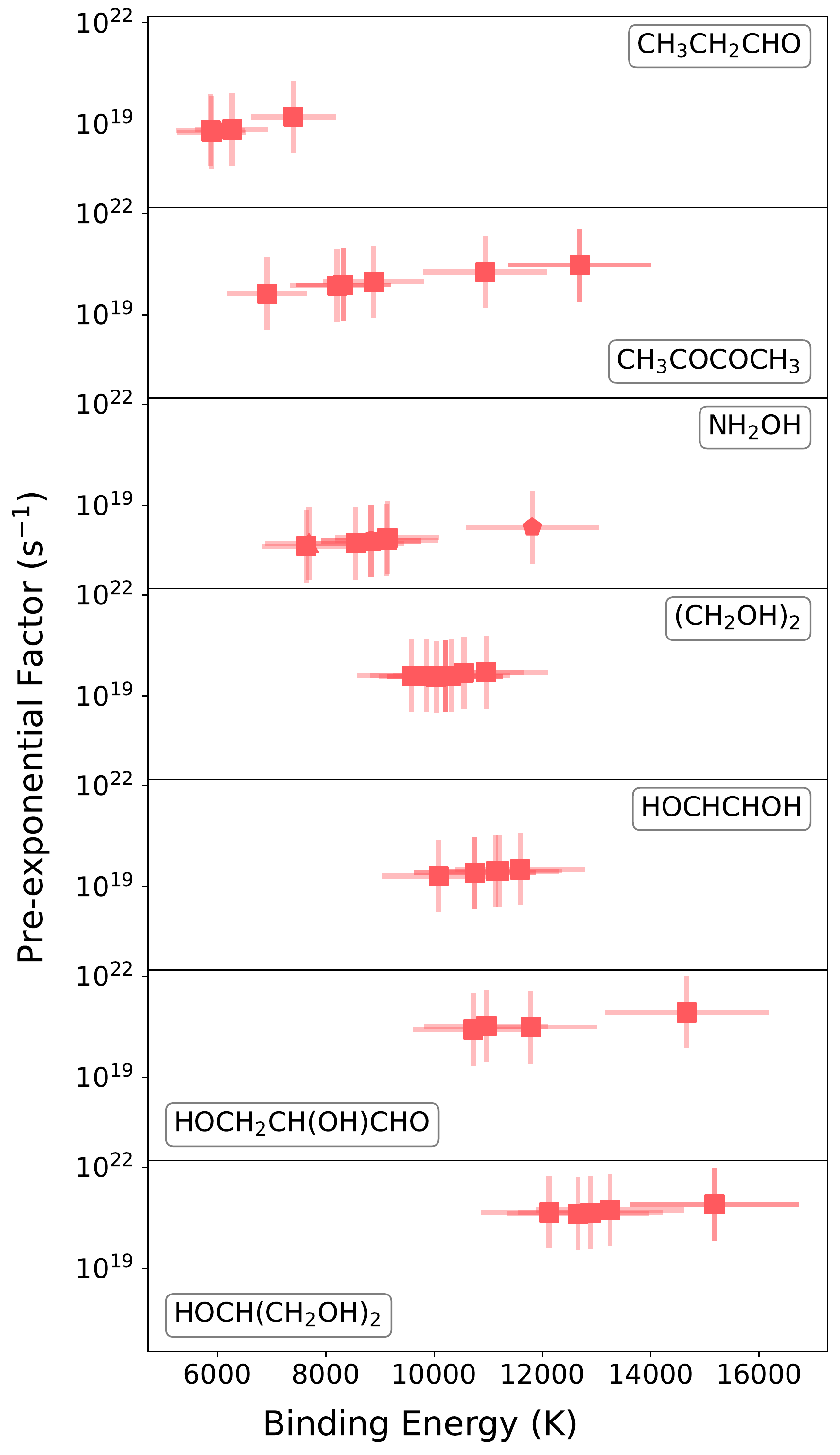}
    \caption{Values of $\nu_{\rm TST}$ and $E_{\rm redhead-TST}$ of all the entries of H$_{3}$CH$_{2}$CHO (propionaldehyde), CH$_{3}$COCOCH$_{3}$ (2,3-Butanedione), NH$_{2}$OH (hydroxylamine), (CH$_{2}$OH)$_{2}$ (ethyleneglycol), HOCHCHOH (1,2-ethenediol), HOCH$_{2}$CH(OH)CHO (glyceraldehyde), and HOCH(CH$_{2}$OH)$_{2}$ (glycerol). Symbols indicate a metallic surface (squares) or carbon surface (pentagon, NH$_{2}$OH panel).}
    \label{fig:cases}
\end{figure}

The Redhead-TST method relies on TPD data, which can show major differences in $T_{\rm peak}$ between experiments. Consequently, this influences the derived $\nu_{\rm TST}$ and $E_{\rm redhead-TST}$ values. Examples of this are shown in Fig. \ref{fig:cases}, where $\nu_{\rm TST}$ and $E_{\rm redhead-TST}$ of all entries of CH$_{3}$CH$_{2}$CHO (propionaldehyde), CH$_{3}$COCOCH$_{3}$ (2,3-Butanedione), NH$_{2}$OH (hydroxylamine), (CH$_{2}$OH)$_{2}$ (ethyleneglycol), HOCHCHOH (1,2-ethenediol), HOCH$_{2}$CH(OH)CHO (glyceraldehyde), and HOCH(CH$_{2}$OH)$_{2}$ (glycerol) are shown. In all cases, marginal variations of at most one order of magnitude are seen in the pre-exponential factors for each molecule. Variations on this level will have a negligible effect on the simulated desorption profiles. However, in several cases large shifts in the retrieved desorption energy are found, with the most extreme scatter seen in CH$_{3}$COCOCH$_{3}$ values, which range from 6920 to 12690~K. These shifts are directly correlated with the peak desorption temperature used as input for the Redhead equation. 

The scatter in peak desorption temperatures, and thus desorption energies, has several explanations. First and foremost the coverage of a species of interest is not known in the majority of cases. Therefore, molecules can span a wide range of (sub)monolayer coverages. At lower coverage, molecules have a tendency to settle in deeper binding sites, which have higher desorption energies. Consequently, there is a correlation between coverage and desorption temperature, where lower coverage results in a shift of desorption to higher temperature \citep[see e.g.,][]{smith2016,he2016b}. This effect is more pronounced on rough surfaces, which have a large range of binding sites, like Amorphous Solid Water (ASW) or the organic residue that is presumably formed in the experiments included in this work. 

The surface on which the experiment is conducted can also play a role. Almost all entries depicted in Fig. \ref{fig:cases} are measured on a metallic surface, but this classification groups many different materials, such as Au, Ag, and Cu, and structure, such as rough, Pt(111), or Mo(110), together. Each metal and surface structure will result in different desorption energies, which in turn shift $T_{\rm peak}$. This shift is exacerbated when other categories of surfaces are used, such as water ice or carbon. For the selected molecules, a clear example is seen for NH$_{2}$OH. One entry is measured on graphite \citep{ioppolo2014} and its retrieved desorption energy is a clear outlier with respect to the other entries of this molecule. Similarly, the type of organic residue formed on a surface during an experiment can also influence the desorption characteristics. This is highlighted by CH$_{3}$COCOCH$_{3}$, for which all entries are measured on the same type of Ag surface, but produced with different irradiated precursors, such as CH$_{3}$CHO, CH$_{3}$CHO:CH$_{3}$COCOCH$_{3}$, and H$_{2}$O:CH$_{3}$CHO. 

Since the desorption parameters of molecules covered in this study have in most cases not been reported, the values listed in this work can be used as an approximation of the pre-factor and desorption energy, for example in chemical models. However, it is important to be aware that the provided values are limited by the quality of the TPD input data.  

\subsection{Redhead-TST method performance}
\label{sec:performance}

\begin{figure}[h]
    \centering
    \includegraphics[width=0.45\textwidth]{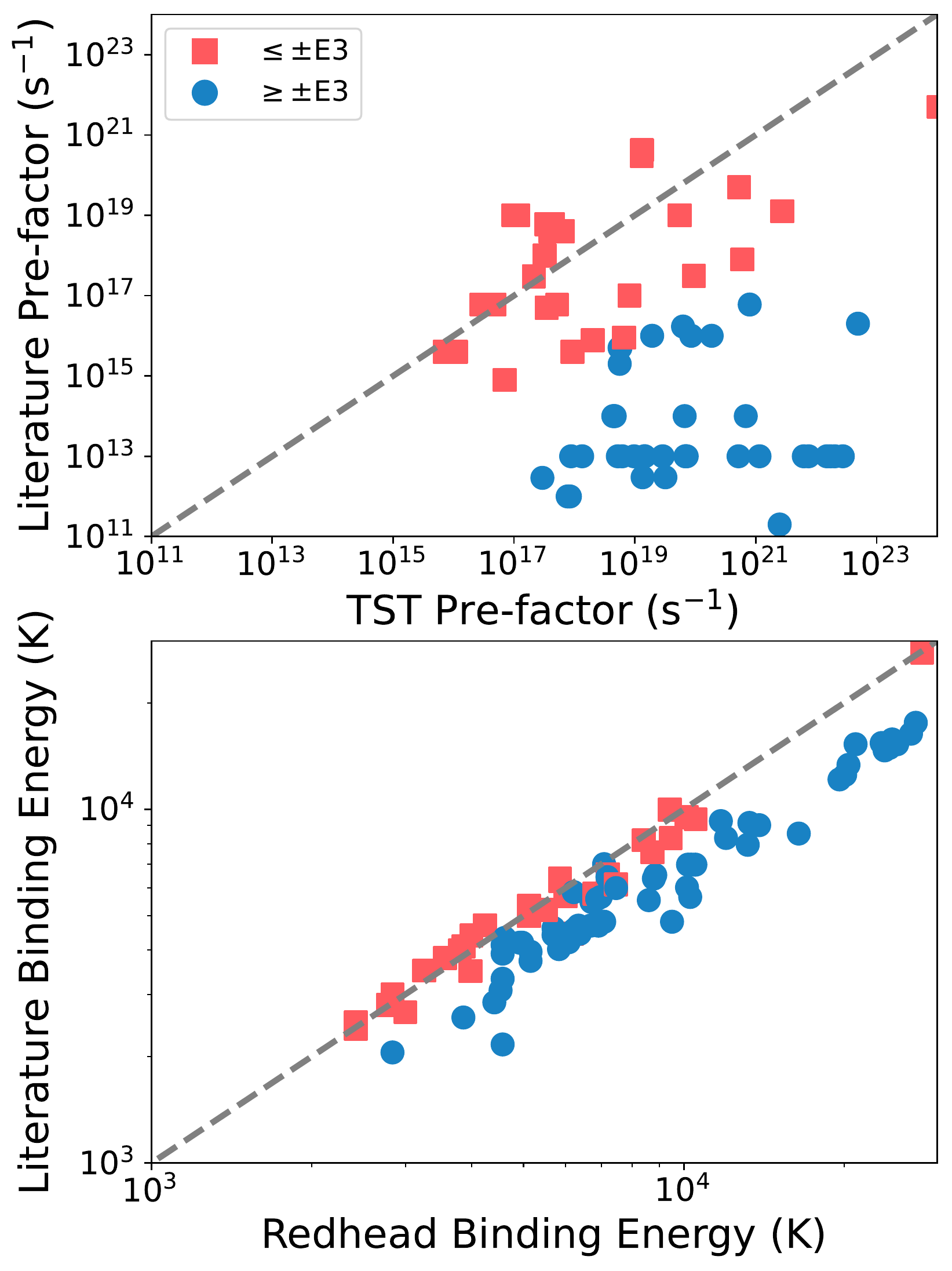}
    \caption{Comparison between $\nu$ (top panel) and $E_{\rm des}$ (bottom panel) values obtained from literature sources, but which have also been analysed with the Redhead-TST method. Red squares indicate entries for which the difference between $\nu_{\rm TST}$ and $\nu_{lit}$ is smaller than three orders of magnitude, whereas the blue circles indicate those with a difference larger than three orders of magnitude.}
    \label{fig:lit_red-tst-comp}
\end{figure}

In Fig. \ref{fig:lit_red-tst-comp} the Redhead-TST and literature $E_{\rm des}$ and $\nu$ values are compared for a number of publications where both these values are available. The top panel of this figure shows the pre-factors values, which are colored based on whether the difference $\Delta$($\nu_{\rm TST}$ - $\nu_{\rm Lit}$) is smaller than $\pm$3 orders of magnitude (red squares) or larger than $\pm$3 orders of magnitude (blue circles). The literature $\nu$ values that are in close agreement with their TST counterpart are found to derive from studies that employ more rigorous methods to determine the pre-factor \citep[e.g.,][]{tait2005a,chaabouni2018,behmard2019,tylinski2020}, although a number of data points are also derived from studies that in fact employ the same TST to calculate $\nu$ \citep{ulbricht2006}. To some extent, the large differences in pre-factors are not surprising, as in many studies $\nu$ is assumed to be 10$^{12}$ or 10$^{13}$ s$^{-1}$. This tendency is visible in the top panel of Fig. \ref{fig:lit_red-tst-comp}. We conclude that the TST method gives accurate approximations of the pre-exponential factor value. 

The subsequent effect of the adopted $\nu$ on the desorption energy is visible in the bottom panel of Fig. \ref{fig:lit_red-tst-comp}. The data points in the bottom panel are again labeled according to the difference in pre-factor values. When the difference in $\nu$ is $\leq$3 decades, the desorption energies are in good agreement and fall on the parity line. Contrary, when there is a large difference between $\nu_{\rm lit}$ and $\nu_{\rm TST}$ there is a prominent difference between $E_{\rm Redhead}$ and $E_{\rm lit}$. Because in all these cases the literature pre-factor is lower than $\nu_{\rm TST}$, the corresponding desorption energy is also lower. This comparison shows that the Redhead-TST method gives accurate desorption parameters that are in good agreement with those derived in experiments, provided these experimental parameters are determined with rigorous laboratory and analysis methods. 

\begin{figure}
    \centering
    \includegraphics[width=0.45\textwidth]{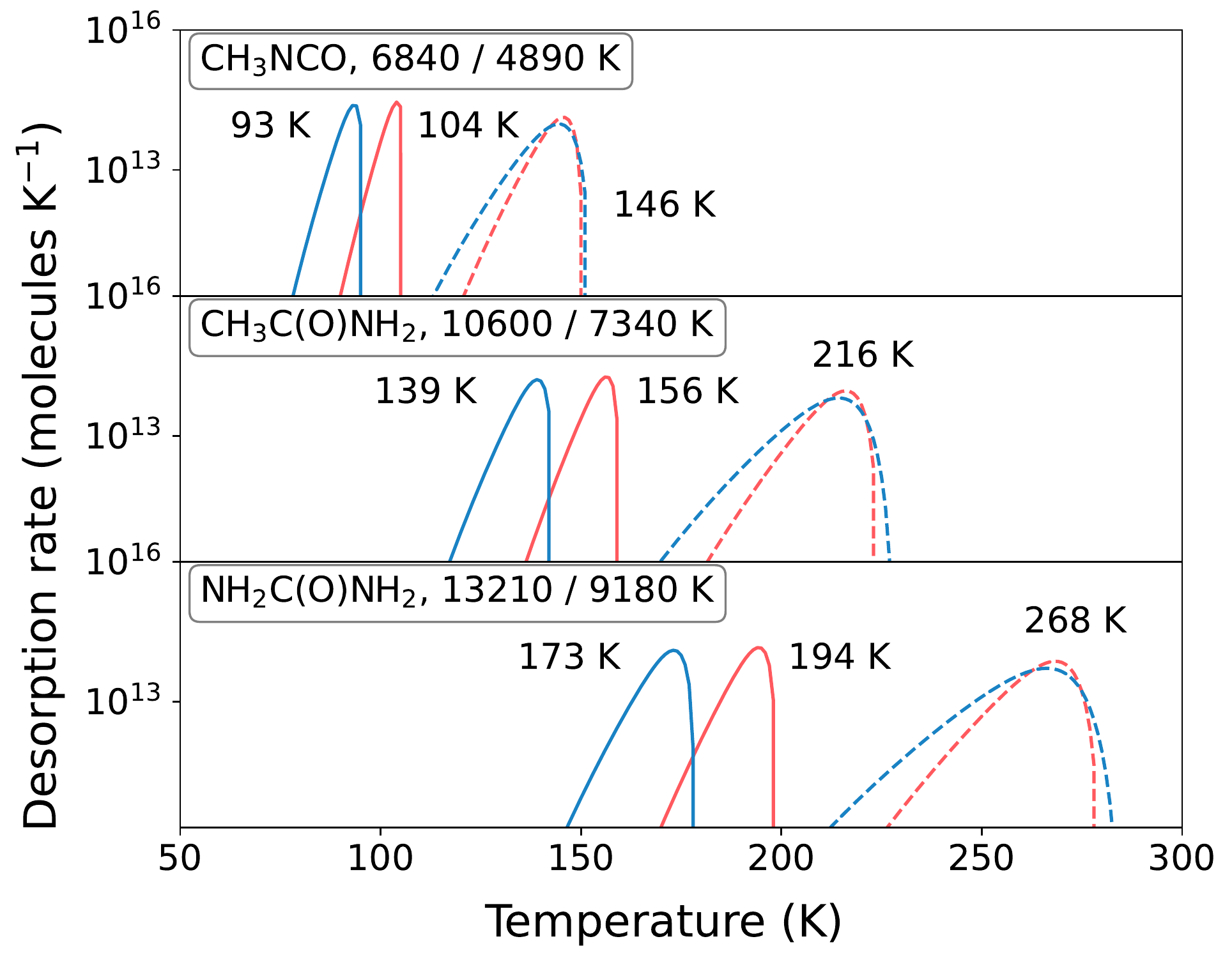}
    \caption{Desorption traces of CH$_{3}$NCO, CH$_{3}$C(O)NH$_{2}$, and NH$_{2}$C(O)NH$_{2}$ with desorption parameters obtained with the Redhead-TST method (red) and the Redhead method, assuming a pre-exponential factor of 10$^{13}$~s$^{-1}$ (blue). Peak desorption temperatures are indicated in the plot for each species. The desorption profile is simulated with a first-order Polanyi-Wigner equation, surface coverage of 1$\times$10$^{15}$ molecules cm$^{-2}$, and heating rate of 1 K century$^{-1}$ (solid lines), which is appropriate for the ISM, or the heating rate of 5 K~min$^{-1}$ (dashed lines) used in these experiments.}
    \label{fig:comp}
\end{figure}

One may argue that the large discrepancies between $E_{\rm Redhead-TST}$ and $E_{\rm lit}$ when $\nu_{\rm lit}$ is assumed will actually not affect the simulated desorption behaviour, as in either case the best fit parameters are generated. An example of this is shown in Fig. \ref{fig:comp}, where data of the molecules CH$_{3}$NCO (methyl isocyanate), CH$_{3}$C(O)NH$_{2}$ (acetamide), and NH$_{2}$C(O)NH$_{2}$ (carbamide) are analysed with the Redhead-TST method (red lines) and the Redhead method, while assuming $\nu$ = 10$^{13}$~s$^{-1}$. The dashed lines show the simulated desorption profiles at a heating rate of 5~K~min$^{-1}$, the same temperature ramp as used in the studies where the TPD data of these molecules are taken from \citep{ligterink2017,ligterink2018a}. The peak desorption temperatures determined with a TST and assumed prefactor are found to be identical and match with the $T_{\rm peak}$ value determined in the laboratory. However, if the heating rate is changed to 1~K~century$^{-1}$, a value that is applicable to interstellar environments, the peak desorption temperatures show a substantial deviation. The simulated profiles with the assumed $\nu$ value, which are lower than the value determined with TST, underestimate the peak desorption temperature by about 10\%. This example highlights the importance of determining desorption parameters, including the pre-exponential factor, as accurately as possible. However, it is important to be aware of the nuances of heating in interstellar environments. Grains can be heated by photons and in the case of very small particles and energetic photons, the grains can be flash heated at rates in the order of K~s$^{-1}$. In these cases desorption energies with assumed pre-exponential factors might be just as applicable, as Fig. \ref{fig:comp} shows. In general, it is recommended to use as accurate a value as possible.

\subsection{Data trends}
\label{sec:trends}

\begin{figure*}[h]
    \centering
    \includegraphics[width=0.90\textwidth]{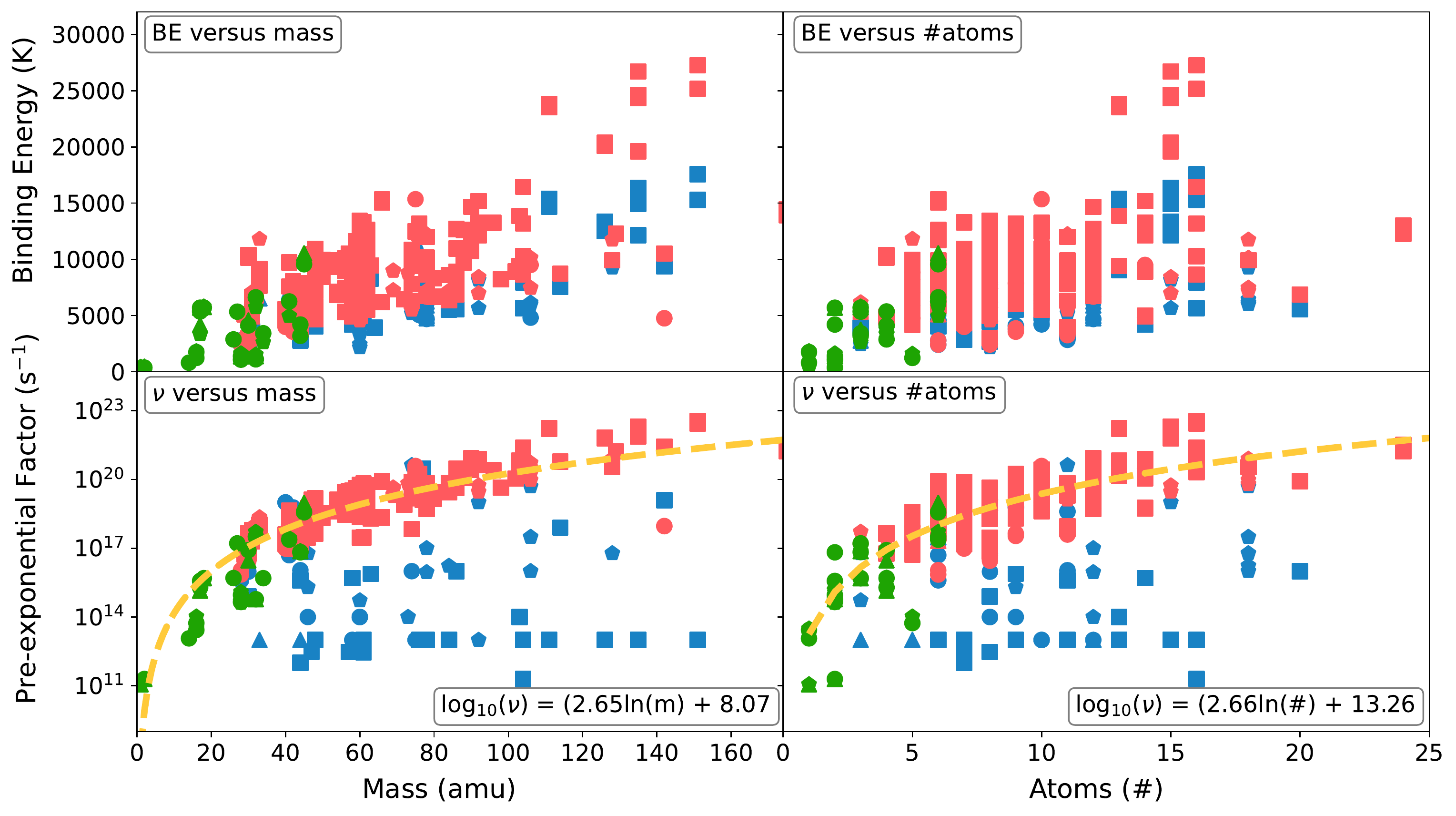}
    \caption{Desorption energy (top row) and pre-exponential factor (bottom row) plotted against molecule mass (left column) and number of atoms (right column). Values derived with the Redhead-TST method are coloured in red, recommended desorption parameters adopted from \citet{minissale2022} in green, and other literature sources are presented in blue. Marker symbol indicates whether a molecule is desorbing from a metallic (squares), water (circles), silicate (triangle), or carbon (pentagon) surface.}
    \label{fig:2}
\end{figure*}

Due to the large amount of collected data, it is possible to investigate trends in pre-factor values and desorption energies. In Fig. \ref{fig:2} $E_{\rm des}$ (top panels) and $\nu$ (bottom panels) are plotted against the molecule mass in amu (left column) and number of atoms of the molecule (right column). Data points obtained with the Redhead-TST method are indicated in red, while those taken from literature are presented in blue. In addition, the recommended desorption parameters of molecules from various surfaces listed in \citet{minissale2022} have been added in green points (data from their Table 2 and 3). Each data point is labelled by its surface category, which includes metallic (squares), carbon (pentagons), silicate (triangles), and water-ice (circles). 

Both Redhead-TST and literature desorption energies show a general tendency to increase with increasing molecule mass or number of atoms. However, most striking is the large spread in desorption energy values for a given molecular mass or number of atoms, which often spans more than 5000~K (top panels Fig. \ref{fig:2}. This spread makes it difficult to retrieve any empirical relationship between the parameters. Even when only data of a specific surface or containing certain functional groups are selected (not shown), no trend or relationship is found. 

The situation is different for the pre-exponential factors. While the laboratory literature data is affected by the tendency to assume $\nu$ = 10$^{13}$ s$^{-1}$, there is a clear trend in the TST data showing a relationship between $\nu_{\rm TST}$ and the molecule mass. These data are fitted with an equation of the form log$_{10}$($\nu$) = a$\cdot$ln($m$) + b, where $m$ is the mass of the molecule in amu (atomic mass units, and the best-fit values are found to be a = 2.65 and of b = 8.07. Because both studied make use of Transition State Theory to determine the pre-factor, the data points of both this work (red) and those of \citet{minissale2022} (green) are used for the fit. The TST data points show a scatter of about one decade around the best-fit line, an uncertainty that has a marginal effect on any simulated desorption profile. The recommended values presented by \citet{minissale2022} are in good agreement with the found equation and show that the equation can also be used for lower mass molecules. This empirical relationship can be used to estimate the pre-exponential factor of a molecule solely based on its mass. This may find use when analysing laboratory data to determine $E_{\rm des}$ values or as an easy way to determine pre-factors for molecules in astrochemical models (see also Sect. \ref{sec:models}). 

\begin{table*}
\caption{\label{tab:averages}Mean desorption parameters.}          
\centering          
\begin{adjustbox}{totalheight=\textheight-2\baselineskip}
\begin{tabular}{lcrr|lcrr}    
\hline\hline    
Molecule & $\nu$ & $E_{\rm des}$ & $T_{\rm peak, ISM}^{a}$ & Molecule & $\nu$ & $E_{\rm des}$ & $T_{\rm peak, ISM}$\\
    & s$^{-1}$ & K & K &  & s$^{-1}$ & K & K\\
\hline
C$_{2}$H$_{4}$ & 9.1e+15 & 2602 & 45 &					CH$_{3}$CHCHOH & 1.6e+19 & 8330 & 125 \\
C$_{2}$H$_{6}$ & 4.8e+16 & 2773 & 46 &					CH$_{3}$OCH$_{2}$OH & 1.9e+19 & 8471 & 127 \\
C$_{3}$H$_{8}$ & 6.5e+17 & 3721 & 59 &					(CH$_{3}$)$_{2}$NCHO & 6.7e+19 & 8785 & 129 \\
CH$_{2}$CHCH$_{3}$ & 3.9e+17 & 3709 & 60 &					{[}NH$_{4}^{+}${]}{[}OCN$^{-}${]} & 2.9e+17 & 8117 & 129 \\
CH$_{3}$PH$_{2}$ & 4.4e+17 & 4662 & 74 &					CH$_{3}$OCH$_{2}$CH$_{3}$ & 2.1e+19 & 8633 & 129 \\
CH$_{3}$I & 9.3e+17 & 4758 & 75 &					C$_{6}$H$_{5}$CCH & 1.1e+20 & 8915 & 130 \\
CH$_{3}$CCH & 1.9e+17 & 4624 & 75 &					CH$_{3}$CH$_{2}$CH$_{2}$OH & 3.8e+19 & 8909 & 132 \\
C$_{4}$H$_{10}$ & 5.7e+18 & 4946 & 76 &					c-C$_{3}$H$_{2}$O & 5.8e+18 & 8766 & 133 \\
CH$_{3}$CHO & 9.7e+17 & 4882 & 77 &					C$_{6}$H$_{5}$CHO & 3.0e+20 & 9502 & 136 \\
H$_{2}$SO$_{2}$ & 2.6e+18 & 5035 & 78 &					CH$_{2}$CH(OH)CH$_{3}$ & 3.3e+19 & 9415 & 139 \\
H$_{2}$CCO & 1.7e+17 & 4895 & 79 &					CH$_{3}$COCOCH$_{3}$ & 1.4e+20 & 9621 & 139 \\
N$_{2}$H$_{2}$ & 5.9e+16 & 4891 & 81 &					NH$_{2}$OH & 1.0e+18 & 8954 & 140 \\
c-C$_{3}$H$_{4}$O & 2.9e+18 & 5319 & 82 &					{[}NH$_{4}^{+}${]}{[}CH$_{3}$COO$^{-}${]} & 1.3e+19 & 9408 & 141 \\
HNCO & 2.9e+17 & 5154 & 83 &					NH$_{2}$CH$_{2}$CN & 2.4e+19 & 9529 & 142 \\
CH$_{2}$NH & 5.3e+16 & 5106 & 84 &					CH$_{3}$CH$_{2}$C$_{6}$H$_{5}$ & 5.3e+20 & 10126 & 144 \\
CH$_{3}$OOCH$_{3}$ & 4.2e+18 & 5560 & 86 &					H$_{2}$PCOOH & 6.4e+19 & 9852 & 144 \\
CH$_{3}$CHCH$_{2}$O & 4.4e+18 & 5715 & 88 &					c-H$_{2}$C$_{3}$O & 1.2e+19 & 9605 & 144 \\
CH$_{3}$CH$_{2}$OCHO & 1.6e+19 & 5833 & 88 &					H$_{2}$POH & 3.0e+18 & 9446 & 145 \\
CH$_{3}$OCHO & 4.7e+18 & 5715 & 88 &					{[}NH$_{4}^{+}${]}{[}HCOO$^{-}${]} & 2.0e+18 & 9426 & 145 \\
CH$_{3}$NC & 2.9e+17 & 5657 & 91 &					NH$_{2}$CH(CH$_{3}$)OH & 4.2e+19 & 9910 & 146 \\
NO$_{2}$ & 3.7e+17 & 5676 & 91 &					C$_{10}$H$_{22}$ & 2.7e+21 & 10517 & 146 \\
HCOPH$_{2}$ & 3.7e+18 & 6006 & 92 &					CH$_{3}$CH$_{2}$CH$_{2}$SH & 9.6e+19 & 10070 & 147 \\
CH$_{3}$CH$_{2}$NH$_{2}$ & 2.3e+18 & 6033 & 94 &					c-NCHCH$_{2}$ & 4.3e+18 & 9733 & 148 \\
P$_{2}$H$_{4}$ & 2.2e+18 & 6183 & 96 &					CH$_{3}$OCOOH & 6.6e+19 & 10151 & 149 \\
CH$_{3}$NH$_{2}$ & 4.9e+17 & 6025 & 96 &					HOCH$_{2}$OH & 1.1e+19 & 9911 & 149 \\
CH$_{3}$CH$_{2}$CHO & 8.8e+18 & 6359 & 97 &					CH$_{2}$CHCH$_{2}$OH & 3.3e+19 & 10139 & 150 \\
HCOSH & 3.8e+18 & 6301 & 97 &					(CH$_{2}$OH)$_{2}$ & 4.2e+19 & 10214 & 151 \\
CH$_{3}$COCH$_{3}$ & 1.0e+19 & 6459 & 98 &					{[}CH$_{3}$NH$_{3}^{+}${]}{[}NCO$^{-}${]} & 6.8e+17 & 9688 & 152 \\
CH$_{3}$CHNH & 1.9e+18 & 6327 & 98 &					HOCH$_{2}$NH$_{2}$ & 1.2e+19 & 10138 & 152 \\
CH$_{2}$CHNH$_{2}$ & 1.9e+18 & 6326 & 98 &					CH$_{2}$C(OH)COOH & 1.5e+20 & 10578 & 153 \\
H$_{3}$SiC(H)CO & 7.9e+18 & 6439 & 98 &					C$_{10}$H$_{8}$ & 5.7e+20 & 10831 & 154 \\
C$_{6}$H$_{14}$ & 8.5e+19 & 6874 & 101 &					HOCH$_{2}$CN & 3.0e+19 & 10375 & 154 \\
CH$_{3}$P$_{2}$H$_{3}$ & 1.4e+19 & 6675 & 101 &					HCOCOOH & 7.8e+19 & 10838 & 158 \\
CH$_{3}$SH & 1.8e+18 & 6522 & 101 &					CH$_{3}$NHCHO & 3.1e+19 & 10708 & 158 \\
CH$_{3}$OCH$_{3}$ & 2.0e+18 & 6503 & 101 &					CH$_{3}$CONH$_{2}$ & 4.0e+19 & 10763 & 159 \\
H$_{2}$S$_{2}$ & 2.8e+18 & 6560 & 101 &					CH$_{3}$COCOOH & 1.9e+20 & 11169 & 161 \\
CH$_{3}$NCO & 6.8e+18 & 6842 & 104 &					H$_{2}$CC(OH)$_{2}$ & 3.4e+19 & 10985 & 162 \\
c-C$_{6}$H$_{12}$ & 6.3e+19 & 7187 & 106 &					(CH$_{2}$-O)x & 4.5e+17 & 10267 & 162 \\
HC$_{2}$CHO & 4.3e+18 & 6957 & 107 &					HOCHCHOH & 2.7e+19 & 10918 & 162 \\
CH$_{2}$CHOH & 2.1e+18 & 6957 & 108 &					c-C$_{6}$H$_{5}$CN & 4.1e+20 & 11628 & 166 \\
N$_{2}$H$_{4}$ & 4.4e+17 & 6779 & 108 &					C$_{6}$H$_{5}$CHCH$_{2}$ & 1.1e+21 & 12136 & 170 \\
CH$_{2}$CHCHO & 8.1e+18 & 7162 & 109 &					HOCH$_{2}$CH(OH)CHO & 4.4e+20 & 12034 & 171 \\
(CHO)$_{2}$ & 6.2e+18 & 7163 & 109 &					C$_{6}$H$_{15}$N$_{3}$ & 1.7e+21 & 12279 & 171 \\
(CH$_{3}$)$_{2}$C$_{6}$H$_{4}$ & 1.4e+20 & 7457 & 109 &					H$_{2}$CO$_{3}$ & 5.3e+19 & 12160 & 178 \\
C$_{4}$H$_{4}$S & 5.0e+19 & 7492 & 111 &					{[}NH$_{4}^{+}${]}{[}NH$_{2}$COO$^{-}${]} & 6.9e+19 & 12166 & 178 \\
HCOOH & 2.0e+18 & 7135 & 111 &					C$_{14}$H$_{10}$ & 3.2e+21 & 13019 & 180 \\
{[}H$_{3}$O$^{+}${]}{[}OCN$^{-}${]} & 3.0e+17 & 7087 & 113 &					HOCH$_{2}$COOH & 1.6e+20 & 12730 & 184 \\
CH$_{3}$CH$_{2}$OH & 5.2e+18 & 7400 & 113 &					CH$_{3}$P(O)(OH)$_{2}$ & 2.5e+20 & 13241 & 190 \\
CH$_{3}$C$_{6}$H$_{5}$ & 4.2e+19 & 7698 & 114 &					HOCH(CH$_{2}$OH)$_{2}$ & 5.7e+20 & 13545 & 192 \\
C$_{6}$H$_{6}$ & 1.2e+19 & 7710 & 116 &					NH$_{2}$COOH & 6.9e+19 & 13276 & 194 \\
CH$_{3}$COOH & 1.3e+19 & 7680 & 116 &					NH$_{2}$CONH$_{2}$ & 3.4e+19 & 13284 & 196 \\
CH$_{3}$C(O)OCH$_{3}$ & 4.1e+19 & 7807 & 116 &					{[}C$_{6}$H1$_{5}$N$_{3}$H$^{+}${]}{[}HCOO$^{-}${]} & 1.7e+21 & 14186 & 198 \\
HOCH$_{2}$CHO & 1.6e+19 & 7797 & 117 &					NH$_{2}$CH$_{2}$COOH & 3.2e+20 & 14390 & 205 \\
HCCNH$_{2}$ & 1.6e+18 & 7587 & 118 &					S$_{3}$ & 2.8e+20 & 14846 & 212 \\
{[}NH$_{4}^{+}${]}{[}CN$^{-}${]} & 1.6e+17 & 7417 & 119 &					S$_{4}$ & 8.3e+20 & 15497 & 218 \\
CHOCH$_{2}$CN & 2.8e+19 & 8099 & 120 &					H$_{2}$P(O)OH & 8.2e+19 & 15207 & 221 \\
CH$_{2}$OCHCN & 2.8e+19 & 8099 & 120 &					CH$_{3}$SCH$_{3}$ & 1.1e+20 & 15350 & 222 \\
CH$_{3}$OOH & 4.2e+18 & 7845 & 120 &					C$_{5}$H$_{6}$N$_{2}$O$_{2}$ & 6.4e+21 & 20232 & 276 \\
(CH$_{2}$NH$_{2}$)$_{2}$ & 1.9e+19 & 8097 & 121 &					C$_{24}$H$_{12}$ & 4.9e+22 & 21013 & 279 \\
CH$_{3}$COCN & 3.3e+19 & 8125 & 121 &					C$_{4}$H$_{5}$N$_{3}$O & 1.7e+22 & 23675 & 318 \\
P$_{3}$H$_{5}$ & 4.4e+19 & 8227 & 122 &					C$_{5}$H$_{5}$N$_{5}$ & 1.5e+22 & 23812 & 320 \\
C$_{8}$H$_{18}$ & 6.0e+20 & 8728 & 124 &					C$_{5}$H$_{5}$N$_{5}$O & 3.2e+22 & 26209 & 348 \\
PH$_{2}$CH$_{2}$PH$_{2}$ & 2.8e+19 & 8307 & 124 &					C$_{32}$H$_{14}$ & 1.0e+24 & 28012 & 356 \\
HCCOH & 1.2e+18 & 8029 & 125 &					C$_{60}$ & 4.1e+24 & 33634 & 399 \\
\hline                  
\multicolumn{8} {c}{\tablefoot{Molecules are sorted by increasing $T_{\rm peak}$ value. $^{a}$The $T_{\rm peak}$ is determined from a desorption profile simulated with a first-order Polanyi-Wigner equation, surface coverage of 1$\times$10$^{15}$ molecules cm$^{-2}$, and heating rate of 1 K century$^{-1}$. Note that these mean values in many cases consists of a superposition of binding environments, many of which are not relevant to interstellar environments (i.e. metallic surfaces). These values should be taken as best available.}}
\end{tabular}
\end{adjustbox}
\end{table*}

\subsection{Desorption parameters and caveats}

For several molecules a large number of entries of the same species have been collected in Table \ref{tab:redhead-tst}. This amount of data can be useful in several cases, such as to analyse the influence of the desorption surface. However, often a single $\nu$ and $E_{\rm des}$ value is desired, for example to be used chemical models. The mean values for $\nu$ and $E_{\rm des}$ for molecules analysed in this work with the Redhead-TST method are presented Table \ref{tab:averages}. 

It is important to note several caveats of the data provided in Table \ref{tab:averages}. First, these desorption parameters are the mean of the available data. Monolayer desorption is a more nuanced process and desorption energies will differ depending on coverage and available binding sites. The mean values presented in this table should only be used when no data is available from more rigorous studies and only for specific applications, such as input for astrochemical models. 

Second, binding environments are grouped together, but one has to be aware that the desorption energies of these molecules will be different when, for example, amorphous water ice or graphite are considered. The most prominent source of desorption parameter data comes from experiments that make use of metallic surfaces (e.g., Au, Ag, Pt substrates). Often the molecule is formed in situ together with other species, and there the binding surface can be seen as a combination of metallic and a refractory organic residue, which makes these environments more realistic. 

In connection to the above points, it is important to note that for some molecules extensive studies have been performed to determine the pure ice (sub)monolayer desorption parameters, including on different surfaces. In particular molecules that are often and abundantly observed in the ISM are on this list, such as CH$_{3}$NH$_{2}$, CH$_{3}$CHO, HCOOH, CH$_{3}$CH$_{2}$OH, CH$_{3}$OCHO, CH$_{3}$COOH, and HOCH$_{2}$CHO \citep[e.g.][]{lattelais2011,bertin2011,burke2015a,burke2015b,burke2015c,chaabouni2018,corazzi2021,ferrero2022}. While these studies are invaluable to assess the influence of different binding environments, in many cases the pre-exponential frequency factor is assumed (e.g., to be 10$^{12}$ or 10$^{13}$~s$^{-1}$), which will affect the determined desorption energy. In Table \ref{tab:redhead-tst} the Redhead-TST analysis of these data are presented when possible. If desorption parameters on surfaces like amorphous or crystalline water, HOPG, or silicates are required, it is recommended to adopt these values. 

While some molecules are well studied, it is also noteworthy that some prominent interstellar molecules have not rigorously been studied in the laboratory, such as dimethyl ether (CH$_{3}$OCH$_{3}$), ethylene glycol ((CH$_{2}$OH)$_{2}$), and acetamide (CH$_{3}$C(O)NH$_{2}$). For some molecules like ethylcyanide (CH$_{3}$CH$_{2}$CN) and vinylcyanide (CH$_{2}$CHCN) only limited data can be found in the literature \citep{toumi2016,kimber2018,couturier-tamburelli2018} and only for the case of multilayer desorption. Because these species are routinely observed in the ISM \citep[e.g.,][]{nazari2022b}, dedicated laboratory studies of these molecules are desired. The parameters of these molecules presented in this study rely solely on chemical TPD data and should be considered the best currently available.

\begin{figure}[h]
    \centering
    \includegraphics[width=0.45\textwidth]{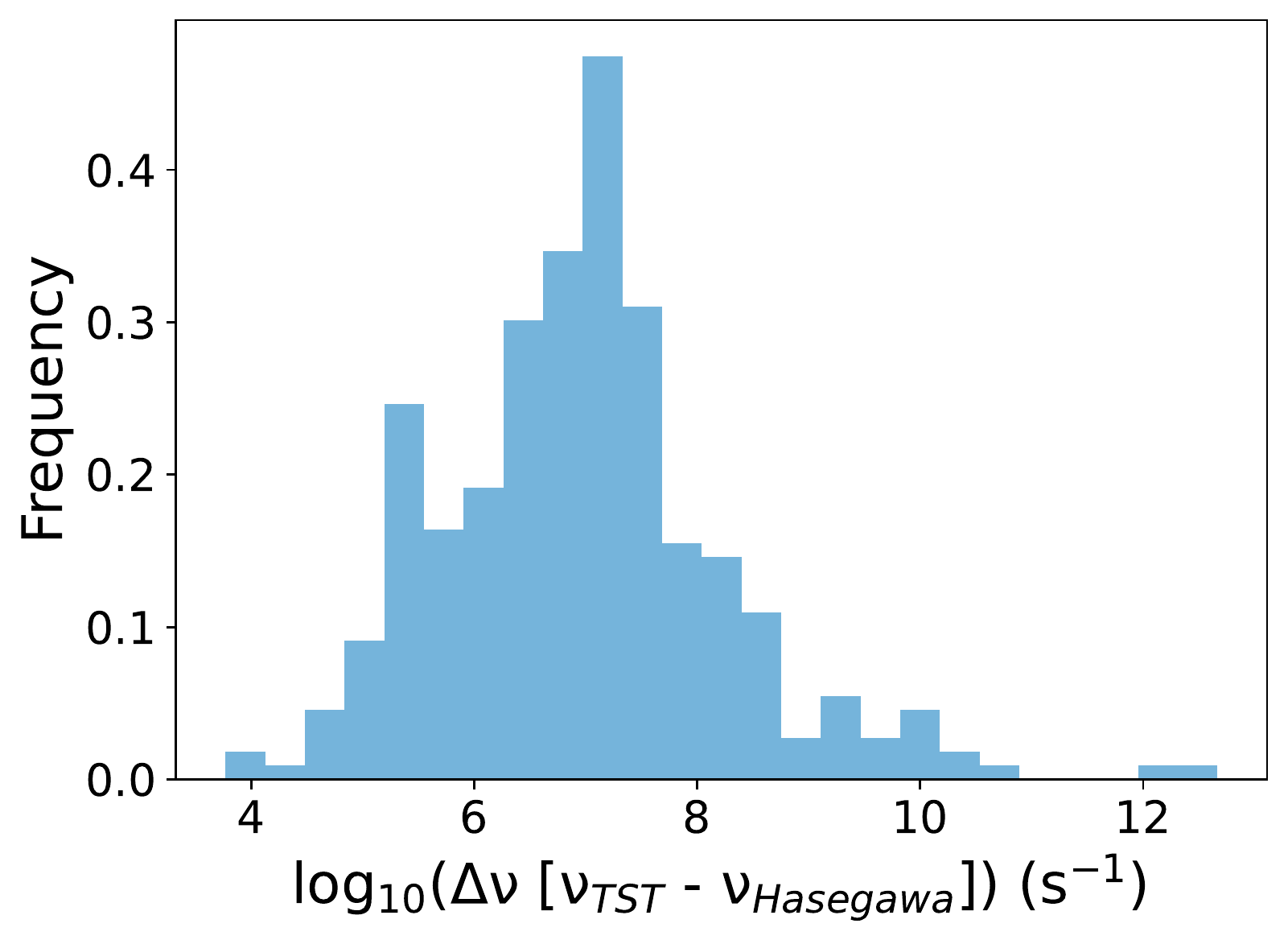}
    \caption{Difference between $\nu_{\rm TST}$ and $\nu_{Hasagawa}$ for molecules presented in this study. Frequencies normalised to one are presented.}
    \label{fig:A_BE_diff}
\end{figure}

\begin{figure*}[h]
    \centering
    \includegraphics[width=0.90\textwidth]{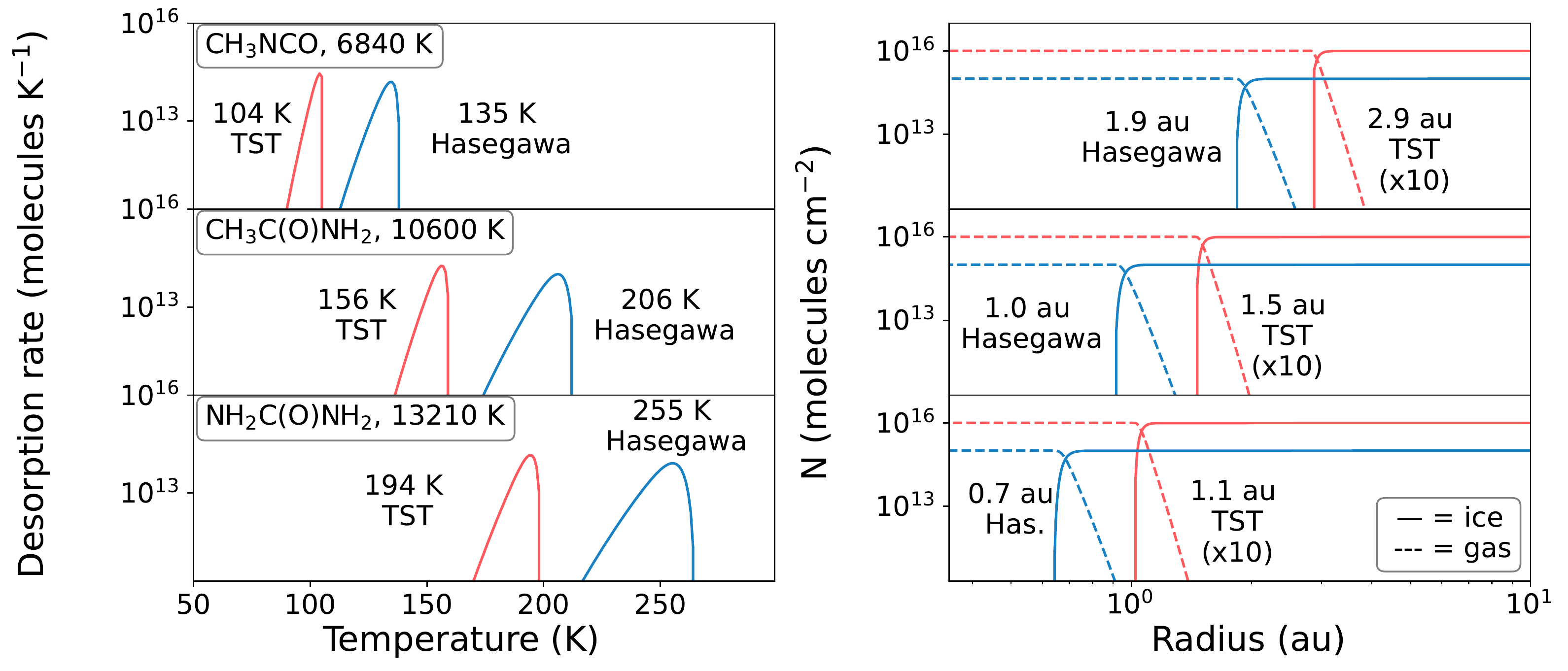}
    \caption{Comparison of desorption profiles generated with the TST (red) or the Hasegawa (blue) prefactor. Left panel: desorption profiles of CH$_{3}$NCO, CH$_{3}$C(O)NH$_{2}$, and NH$_{2}$C(O)NH$_{2}$. The desorption energies are determined with the Redhead-TST method using data from \citet{ligterink2017} and \citet{ligterink2018a} (see Table \ref{tab:redhead-tst} and top left corners). Desorption profiles are plotted for $\nu_{\rm TST}$ (red) and with $\nu_{\rm Hasegawa}$ (blue). The desorption profile is simulated with a first-order Polanyi-Wigner equation, surface coverage of 1$\times$10$^{15}$ molecules cm$^{-2}$, and heating rate of 1 K century$^{-1}$. Peak desorption temperatures decrease by $\sim$30\% when realistic pre-factors are used.
   Right panel: ice (solid) and gas (dashed) abundances plotted against radius (au) of a protoplanetary disk for $\nu_{\rm TST}$ (red) and with $\nu_{\rm Hasegawa}$ (blue). An average disk temperature profile of $T(r) = 200 \times (r / 1 AU)^{0.62}$ K is used \citep{andrews2007}. For each 1~K temperature step, the number of desorbed molecules is determined, which summed give the gas-phase column density and subtracted from a starting value of 1$\times$10$^{15}$~cm$^{-2}$ give the ice column density. The TST abundances are offset by a factor of 10 for easier viewing. The peak radii of the ice-gas inversion are indicated and are shown to shift outward by $\sim$50\% when realistic pre-factors are used.}
    \label{fig:par_var}
\end{figure*}

\section{Astrophysical implications}
\label{sec:implications}

\subsection{Impact of the pre-factor on astrochemical models}
\label{sec:models}

Astrochemical models make use of the Polanyi-Wigner equation or a variation thereof to calculate the desorption rates of molecules on dust grains \citep[e.g.,][]{kulterer2020}. While for these models desorption energies are generally taken from experimental and theoretical literature, the $\nu$ value is calculated with an empirical formula, often the harmonic oscillator relation presented in \citet{hasegawa1992}:

\begin{equation}
    \nu = \sqrt{2 k_{\rm B} n_{\rm ss} E_{\rm des} / \pi^{2} m},
    \label{eq:balance}
\end{equation}

which uses the molecule mass $m$, the molecule binding energy $E_{\rm des}$, and the number of binding sites per grain surface area $n_{\rm ss}$ as input to calculate the pre-factor. Except for small molecules and atoms, this equation underestimates the value of $\nu$, often by multiple decades \citep{minissale2022}. This is also found for molecules considered in this study. Figure \ref{fig:A_BE_diff} shows the difference between $\nu_{\rm TST}$ and $\nu_{\rm Hasegawa}$ and visualises that for the majority of cases $\nu_{\rm TST}$ is at least seven decades larger than the pre-factor calculated with the Hasegawa equation.

The difference in adopted pre-factor can have a strong impact on astrochemical models. Combining a desorption energy that is determined with Redhead-TST or any other accurate method with a pre-factor determined by the Hasagawa equation can severely misrepresent the desorption behaviour of the molecule \citep[see Fig. 2 in][]{ceccarelli2022}. This is exemplified in Fig. \ref{fig:par_var} for the molecules CH$_{3}$NCO, CH$_{3}$C(O)NH$_{2}$, and NH$_{2}$C(O)NH$_{2}$. For these species $E_{\rm Redhead-TST}$ is determined with data from \citet{ligterink2017} and \citet{ligterink2018a} and subsequently plotted with $\nu_{\rm TST}$ (red) and $\nu_{\rm Hasegawa}$ (blue). The desorption profiles (left panel) plotted with $\nu_{Hasegawa}$ are all shifted to higher temperatures and misrepresent the actual desorption profile. For these examples the peak desorption temperature is about 30\% higher than what it should be. In the right panel the gas and ice column densities are plotted against the radius of a protoplanetary disk, which has the average temperature profile $T(r) = 200 \times (r/ 1 AU)^{0.62}$ K, as found by \citet{andrews2007}. For each 1~K temperature step, the number of desorbed molecules is determined, which summed give the gas-phase column density and subtracted from a starting value of 1$\times$10$^{15}$~cm$^{-2}$ give the ice column density. As expected from their peak desorption temperatures, the ice-gas inversion point lies at a larger radius when the accurate TST pre-factor is used. Using $\nu_{\rm Hasegawa}$ results in the inversion point shifting $\sim$50\% inwards. 

Due to its impact on astrochemical models, it is recommended to use more accurate pre-exponential factor values. One way is to implement the empirical equation presented in Sect. \ref{sec:trends}, with the caveat that this equation is ill suited to determine pre-factors for atoms and diatomic molecules. Better yet is to use the $\nu$ and $E_{\rm des}$ that belong together as presented in the literature source as direct input for the chemical model, to prevent any misrepresentation.

\subsection{Peak desorption temperatures}

For simplicity and consistency, desorption profiles and peak desorption temperatures are calculated using Eq. \ref{eq:polwig} and a heating rate of 1~K~century$^{-1}$ in this work. However, this heating rate is not appropriate for all interstellar environments and peak desorption temperatures are often determined with other equations, such as the adsorption-thermal-desorption balance presented by \citet{hasegawa1992} \citep[see also][]{nazari2021}:

\begin{equation}
    \frac{n_{\rm ice}}{n_{\rm gas}} = \frac{\pi a^{2}_{\rm grain} n_{\rm grain} \sqrt{3 k_{\rm B} T m^{-1}}}{e^{-E_{\rm des}/T} \sqrt{2 k_{\rm B} n_{\rm ss} E_{\rm des} \pi^{-2} m^{-1}}}.
    \label{eq:balance1}
\end{equation}

As input, this equation takes the grain size $a_{\rm grain}$ (set to 0.1 $\mu$m), the grain number density $n_{\rm grain}$ (set to 1.0$\times$10$^{-12}$ $n_{\rm H}$, with $n_{\rm H}$ the gas density), the sticking efficiency $S$ (set to 1), the gas and grain temperature $T$, the molecule mass $m$, the molecule binding energy $E_{\rm des}$, and the number of binding sites per grain surface area $n_{\rm ss}$ to calculate the ice-to-gas ratio. The peak desorption temperature is taken as the point where ice and gas molecular abundances are equal, that is $n_{\rm ice}$ / $n_{\rm gas}$ = 1. Instead of the peak desorption temperature depending on the heating rate values, in this equation it primarily depends on the gas density of the environment. Furthermore, we note that this equation in its original form makes use of the harmonic oscillator relation to calculate the pre-exponential frequency in the denominator. As discussed Sect. \ref{sec:models}, this equation gives inaccurate values for larger molecules and therefore the equation should rather take accurate pre-factor into account:

\begin{equation}
    \frac{n_{\rm ice}}{n_{\rm gas}} = \frac{\pi a^{2}_{\rm grain} n_{\rm grain} \sqrt{3 k_{\rm B} T m^{-1}}}{\nu_{\rm TST} e^{-E_{\rm des}/T}}.
    \label{eq:balance2}
\end{equation}

Using the latter equation, the peak desorption temperatures where calculated for $n_{\rm H}$ = 1$\times$10$^{7}$ cm$^{-3}$ (appropriate for molecular clouds) and $n_{\rm H}$ = 1$\times$10$^{12}$ cm$^{-3}$ (appropriate for protoplanetary disk environments) and compared with the $T_{\rm peak}$ determined from the peak of the Polanyi-Wigner desorption profile using a heating rate of $\beta$ = 1~K~century$^{-1}$. Desorption parameters listed in Table \ref{tab:averages} were used. In Table \ref{tab:balance} the results for a selection of twenty molecules are presented. 

\begin{table}
\caption{\label{tab:balance}Comparison peak desorption temperatures.}          
\centering          
\begin{tabular}{lrrrrr}    
\hline\hline    
Molecule & $T_{\rm p, PW}$ & $T_{\rm p, ATD}$ & Diff. & $T_{\rm p, ATD}$ & Diff. \\
& & $n_{\rm H, 7}$ & & $n_{\rm H, 12}$ & \\
& K & K & \% & K & \% \\
\hline
C$_{2}$H$_{4}$ & 45 & 43 & 96 & 54 & 119 \\
CH$_{3}$CHO & 77 & 75 & 98 & 92 & 119 \\
H$_{2}$CCO & 79 & 78 & 98 & 95 & 120 \\
CH$_{2}$NH & 84 & 83 & 99 & 102 & 121 \\
CH$_{3}$SH & 101 & 100 & 99 & 121 & 120 \\
CH$_{3}$OCH$_{3}$ & 101 & 99 & 98 & 121 & 120 \\
CH$_{3}$NCO & 104 & 102 & 98 & 124 & 119 \\
HC$_{2}$CHO & 107 & 105 & 98 & 127 & 119 \\
CH$_{2}$CHOH & 108 & 106 & 98 & 129 & 120 \\
(CHO)$_{2}$ & 109 & 107 & 99 & 130 & 119 \\
HCOOH & 111 & 109 & 98 & 133 & 119 \\
CH$_{3}$CH$_{2}$OH & 113 & 111 & 99 & 135 & 120 \\
CH$_{3}$COOH & 116 & 114 & 98 & 138 & 119 \\
NH$_{2}$OH & 140 & 139 & 99 & 169 & 121 \\
H$_{2}$PCOOH & 144 & 143 & 99 & 172 & 119 \\
HOCH$_{2}$CN & 154 & 152 & 99 & 184 & 119 \\
HCOCOOH & 158 & 157 & 99 & 188 & 119 \\
NH$_{2}$CONH$_{2}$ & 196 & 195 & 99 & 235 & 120 \\
S$_{4}$ & 218 & 216 & 99 & 258 & 118 \\
C$_{60}$ & 399 & 417 & 104 & 487 & 122 \\
\hline                  
\end{tabular}
\tablefoot{Peak desorption temperatures for various molecules, using the Polanyi-Wigner equation ($T_{\rm p, PW}$, with $\beta$ = 1~K~century$^{-1}$), and the adsorption-thermal-desorption equation ($T_{\rm p, ATD}$), for $n_{\rm H}$ = 1$\times$10$^{7}$~cm$^{-3}$ ($n_{\rm H, 7}$) and $n_{\rm H}$ = 1$\times$10$^{12}$~cm$^{-3}$ ($n_{\rm H, 12}$). The differences between the adsorption-thermal-desorption and Polanyi-Wigner peak desorption temperatures are indicated in percent of ($T_{\rm p, ATD}$ / $T_{\rm p, PW}$). Desorption parameters from Table \ref{tab:averages} are used.}
\end{table}

The adsorption-thermal-desorption equation a low density ($n_{\rm H}$ = 1$\times$10$^{7}$~cm$^{-3}$) shows a small difference with the Polanyi-Wigner results. One aspect that affects this comparison is the definition of the peak desorption temperature of the adsorption-thermal-desorption balance equation, which is located at the point where $n_{\rm ice}$ / $n_{\rm gas}$ = 1. However, for the $T_{\rm peak}$ from the Polanyi-Wigner equation this is usually $n_{\rm ice}$ / $n_{\rm gas} \leq$ 0.1. Correcting for this discrepancy raises the adsorption-thermal-desorption balance peak desorption temperatures by several K. In turn, this make $T_{\rm peak, PW}$ and $T_{\rm peak, ATD}$ for the low-density scenario virtually identical. Quite different is the situation for the high-density ($n_{\rm H}$ = 1$\times$10$^{12}$~cm$^{-3}$) results of the adsorption-thermal-desorption balance equation, which show peak desorption temperatures that are approximately 20\% higher than those obtained with the Polanyi-Wigner equation. The peak desorption temperatures listed in this study might therefore not be representative for all interstellar environments and source-specific modelling is required to accurately determine desorption fronts.

\subsection{Salt desorption}

\begin{figure}[h]
    \centering
    \includegraphics[width=0.45\textwidth]{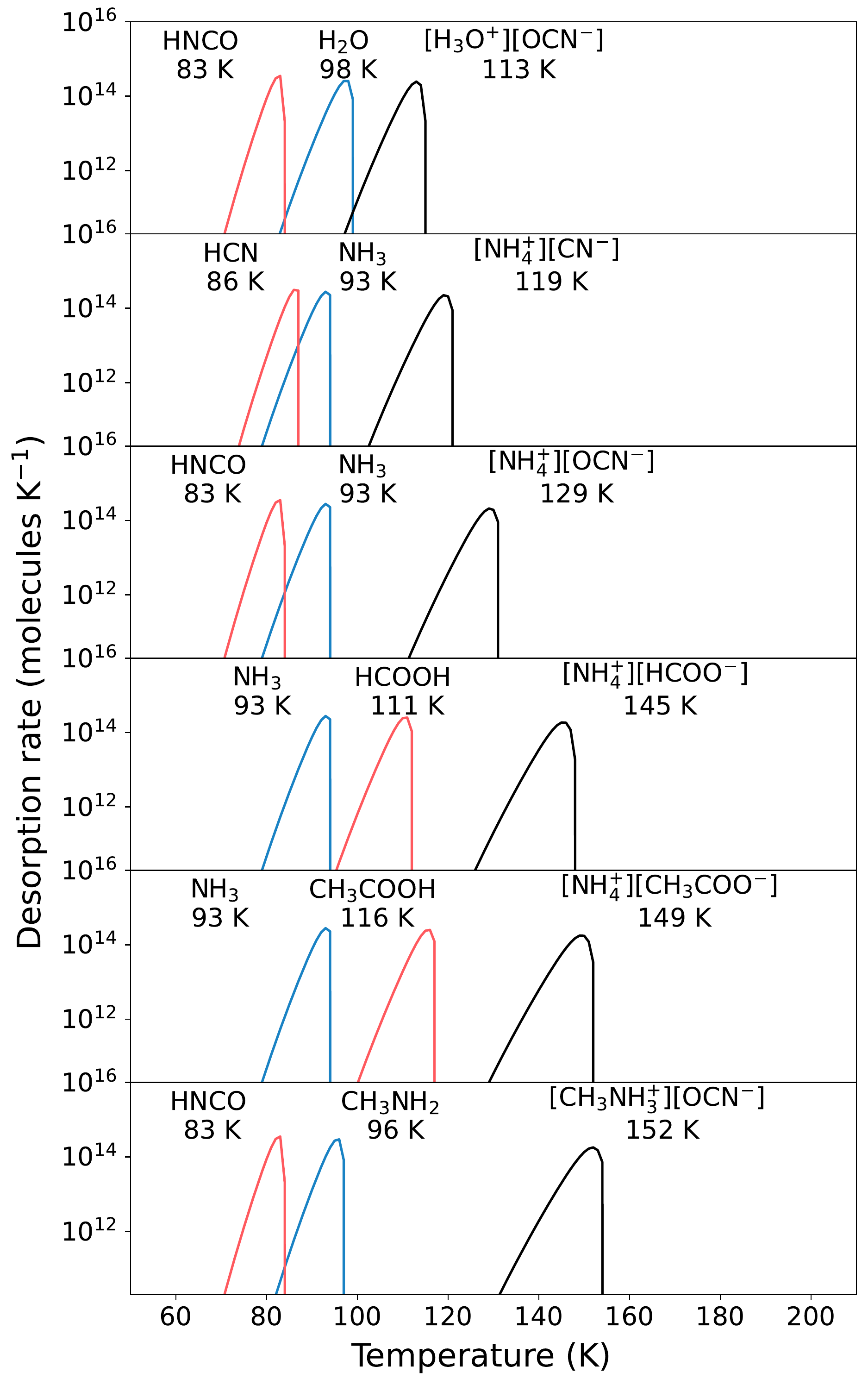}
    \caption{Desorption profiles of acids (red), bases (blue), and the resulting salt (black). Desorption energies and pre-factors for H$_{2}$O, NH$_{3}$, and HCN are obtained from \citet{minissale2022}, while the parameters for the remaining molecules are taken from Table \ref{tab:averages}. Peak desorption temperatures are indicated in the plot for each species. The desorption profile is simulated with a first-order Polanyi-Wigner equation, surface coverage of 10$^{15}$ molecules cm$^{-2}$, and heating rate of 1 K century$^{-1}$.}
    \label{fig:salt_des}
\end{figure}

Several molecules detected in the interstellar medium can be classified as acids ([AH], e.g., HCOOH, HNCO, CH$_{3}$COOH) or bases ([B], e.g., NH$_{3}$, CH$_{3}$NH$_{2}$). In ice mantles, these species can engage in acid-base reactions and form organic salts via the reaction [AH] + [B] $\rightarrow$ [A$^{-}$][BH$^{+}$]. Several of such salts are included in this study and their desorption energies are determined. The pre-factors of these salts are determined by adding the pre-exponential factor values of its individual components ([AH] and [B]), see Sect. \ref{sec:redhead-tst}. While there is a chemical diversity in the salts, they often include ammonia as the base. In the interstellar medium, ammonia is also expected to be a prominent base, due to its high ice abundance \citep[up to 10\% w.r.t. H$_{2}$O,][]{boogert2015}.  In all cases, $E_{\rm des}$ of the salt is larger than the value of its individual components and therefore the salts will reside on dust grains at higher temperatures \citep[see also e.g.,][]{kruczkiewicz2021}. This is exemplified in Fig. \ref{fig:salt_des} where the desorption profiles of various salts and their corresponding acid and base are plotted. A first-order Polanyi-Wigner equation is used, with a monolayer coverage of 10$^{15}$  molecules cm$^{-2}$ and a heating rate of 1~K~century$^{-1}$. Differences in peak desorption temperatures range from just over 10~K for the H$_{2}$O--HNCO system, to more than 60~K for the CH$_{3}$NH$_{2}$--HNCO system. 

The ability to lock up molecules in the form of salts can have number of implications. Large quantities of organic molecules with amino (-NH$_{2}$) or carboxylic acid (-COOH) groups can be trapped in organic salt complexes and remain on dust grains and larger bodies at elevated temperatures. Recent analysis of samples collected and return from the astroid Ryugu show that it contains high concentrations of amines (e.g., CH$_{3}$NH$_{2}$) and acids \citep[HCOOH and CH$_{3}$COOH,][]{naraoka2023}. Because these molecules are found to not be trapped in minerals or other organic matter, the authors suggest that these species reside in the material as salts in order to explain why these volatiles are still present. Our results underline this conclusion, although questions remains if the salts included in this study, which are still relatively volatile, could have survived the heating ($\geq$~300~K) and hydrothermal stages during the formation of Ryugu \citep{nakamura2022}. Further desorption studies of salts complexes could help unravel in which salt form amines and acids are locked up in asteroid Ryugu.

Organic salt complexes also provide a molecular reservoir that remains available for grain surface chemistry for a longer time (or rather to higher temperatures) but not for gas-phase reactions. Ammonia is the dominant nitrogen carrier in observed interstellar ice, but it is also relatively volatile. Locking ammonia up in ammonium salts (NH$_{4}^{+}$) could mean that a substantial atomic nitrogen reservoir is available on interstellar grains long after NH$_{3}$ or H$_{2}$O have desorbed, see Fig. \ref{fig:salt_des}. Laboratory investigations of the processing of salts with energetic radiation at elevated temperatures ($\geq$100~K) and in water-free environments, are relevant avenues to investigate the formation of prebiotic molecules \citep[see for example][]{bossa2009}.

As salts desorb at higher temperatures, they can give misleading indications of sublimation fronts as determined with observations of interstellar environments. For example, \citet{lee2022} measure the spatial distribution of various molecules toward the HH212 protoplanetary disk and find similar radial distributions of HNCO and NH$_{2}$CHO. If the presence of these molecules in the gas can entirely be explained by ice desorption, this co-spatial distribution is peculiar as these molecules have significantly different desorption parameters of 2.9$\times$10$^{17}$~s$^{-1}$ and 5154~K (this work) and 3.69$\times$10$^{18}$~s$^{-1}$ and 9561~K \citep{minissale2022} for HNCO and NH$_{2}$CHO, respectively. One would expect that HNCO desorbs at a much lower temperature or larger radius than NH$_{2}$CHO. However, if instead most of the HNCO available in the ice has reacted with NH$_{3}$, it would be locked up in the [NH$_{4}^{+}$][OCN$^{-}$] salt. This salt has the desorption parameters 2.9$\times$10$^{17}$~s$^{-1}$ and 8117~K, which are close to those of formamide. The salt can therefore explain the co-spatial distribution of HNCO and NH$_{2}$CHO due to thermal desorption of the [NH$_{4}^{+}$][OCN$^{-}$] salts, which upon desorption dissociates into its individual components, HNCO and NH$_{3}$ \citep[see also][]{lee2022}.

Thus far two molecules have been detected in the interstellar medium that are presumed to be part of salt complexes. These are the cyanate anion (OCN$^{-}$) and the ammonium cation NH$_{4}^{+}$ \citep{lacy1984,keane2001b,pontoppidan2003,vanbroekhuizen2005}. Recently, these species have also been detected in observations with the new \textit{James\ Webb\ Space\ Telescope} \citep[JWST,][]{mcclure2023}. The unprecedented sensitivity of JWST opens up two avenues of research. First, it is possible to look for other salt components at lower abundance or with weaker spectroscopic features, for example the cyanide anion (CN$^{-}$) and carboxylate anions (R-COO$^{-}$). Second, searches for salts in water poor or free interstellar environments, for example molecular clouds that interact with the warm gas of a protostellar outflow or the edges of photo dominated regions (PDRs), can be conducted. Because the bulk ice species have been removed from the grains in these regions, the less abundant salt and organic species are easier to detect. Furthermore, the presence and shape of salt spectroscopic signatures will provide information about the chemical and physical history of the environment. 

\subsection{Chemical and elemental composition above the water-snow line}
\label{sec:element}

\begin{figure*}[h]
    \centering
    \includegraphics[width=0.90\textwidth]{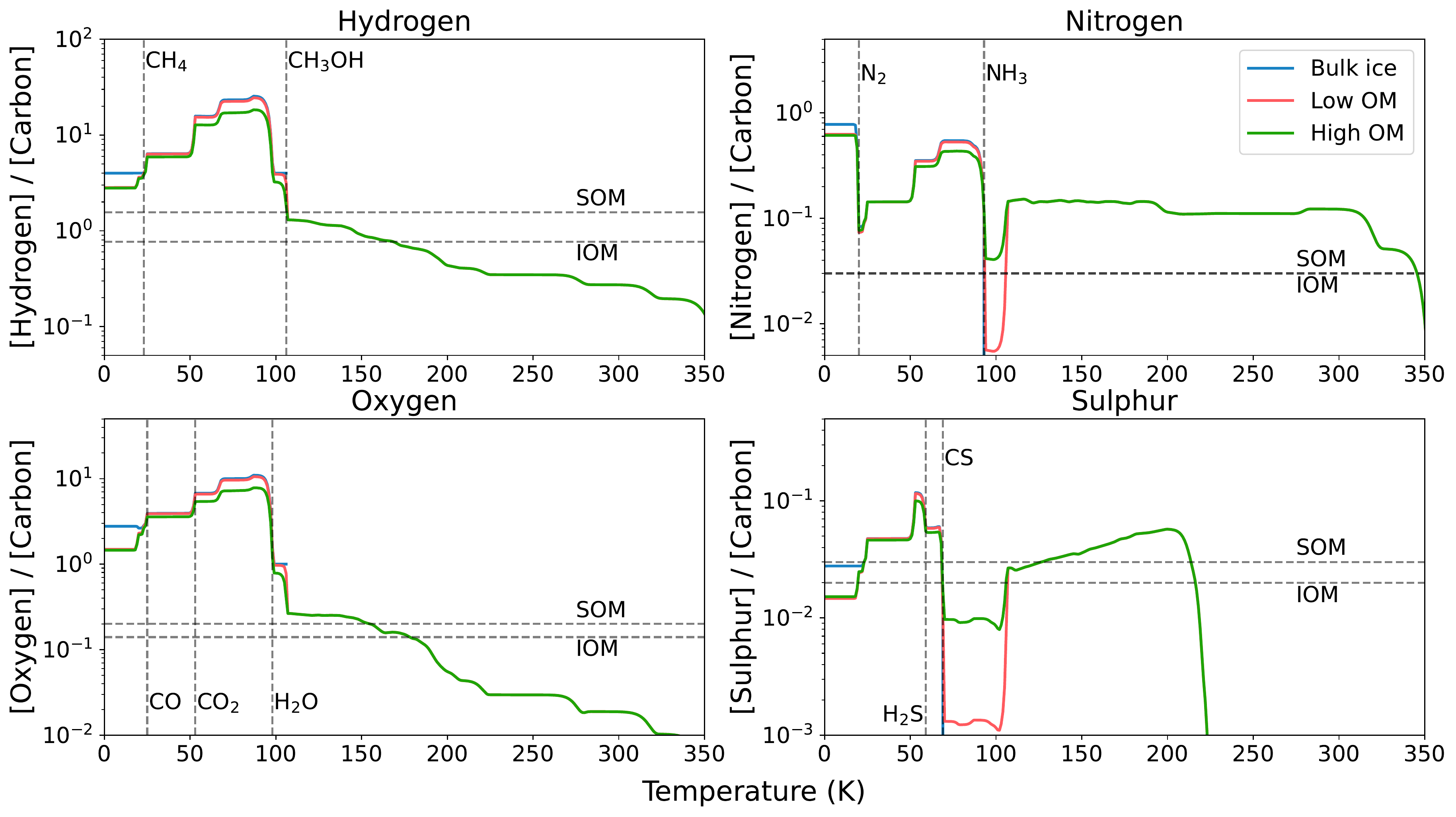}
    \caption{Ice elemental composition as [X]/[Carbon] for Hydrogen, Nitrogen, Oxygen, and Sulphur as a function of temperature. Peak desorption temperatures for selected species are indicated in the plot (vertical dashed lines). Elemental compositions of Soluble Organic Matter (SOM) obtained from the Murchison meteorite \citep{schmitt-kopplin2010} and averaged Insoluble Organic Matter \citep[IOM,][]{alexander2017} are indicated (horizontal dashed lines). The desorption profiles are simulated with a first-order Polanyi-Wigner equation, variable surface coverages, and a heating rate of 1 K century$^{-1}$. Low and high OM represent low and high fraction of organic molecules, respectively. }
    \label{fig:elem_rat}
\end{figure*}

\begin{figure}[h]
    \centering
    \includegraphics[width=0.45\textwidth]{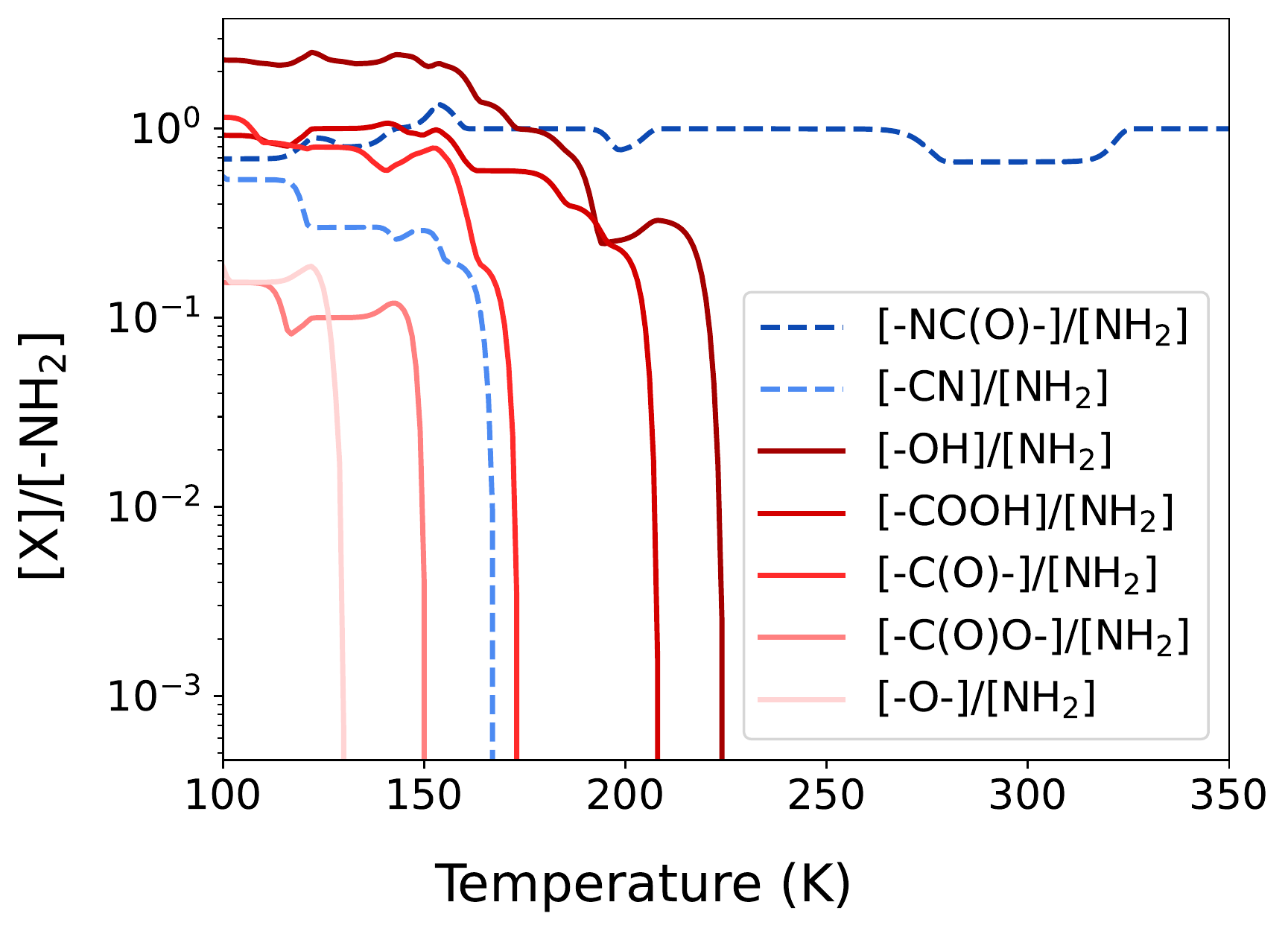}
    \caption{Functional chemical group composition [X]/[-NH$_{2}$] for oxygen-bearing groups -OH, -COOH, -C(O)-, -C(O)O-, and -O- (solid red lines) and nitrogen-bearing groups -NC(O)- and -CN (dashed blue lines) as a function of temperature. The desorption profiles are simulated with a first-order Polanyi-Wigner equation, variable surface coverages, and a heating rate of 1 K century$^{-1}$.}
    \label{fig:functional_groups}
\end{figure}

Within star- and planet-forming environments temperature gradients and the associated desorption fronts of molecules, also called snow lines, play an important role in forming planetary objects and setting their chemical and elemental composition. Examples of such lines are the nitrogen and water snowline or the soot line, which is driven by the sublimation of large carbon-dominated molecules such as polycyclic aromatic hydrocarbons (PAHs). These lines are invoked to explain the high atomic nitrogen abundance in Jupiter or the comparatively low carbon content of Earth \citep{bosman2019,oberg2021,li2021carbon}. Because of their high abundances, observational, modelling, and experimental efforts have focused on the main ice species to investigate desorption fronts. Since most organic molecules except for CH$_{3}$OH and CH$_{4}$ are found at low abundances, these species will not affect the elemental composition of a protoplanetary disk or create snow lines that are observable. However, the combined inventory of organic molecules may affect the elemental composition. \textit{ISO} and \textit{Spitzer} IR observation of various interstellar sources have found significant features in the 5--8~$\mu$m region \citep{gibb2002,boogert2008}. These features have at least in part been assigned to an organic residue consisting of a variety of molecules, including HCOOH and HCOO$^{-}$. Therefore, the data presented in this paper are used in this section to investigate how a large reservoir of diverse organic molecules affects elemental ratios and the chemical composition, for example in a protoplanetary disk. 

Elemental ratios of hydrogen, nitrogen, oxygen, phosphorus, and sulphur over carbon are determined for a molecular inventory that consists of the species listed in this work and desorption parameters presented in Table \ref{tab:averages}, supplemented by bulk ice species with parameters taken from Table 3 of \citet{minissale2022}. The ice abundances are set with respect to water. The following fractions are used: CO, N$_{2}$, and CO$_{2}$ at 0.25, CH$_{3}$OH at 0.1, O$_{2}$, CH$_{4}$, NH$_{3}$, and H$_{2}$CO at 0.05, and H$_{2}$S, CS, and HCN at 0.01. Combined, the aforementioned species are indicated as the bulk ice species. All remaining organic molecules listed in this work, as well as C$_{2}$H$_{2}$, CH$_{3}$CN, NH$_{2}$CHO taken from \citet{minissale2022} are included in a low and high fraction scenario where each organic molecule contributes at 10$^{-4}$ and 10$^{-3}$ level, respectively. The desorption profile, including ice and gas abundances are simulated with a first-order Polanyi-Wigner equation, variable abundances according to fractions with respect to water, and a heating rate of 1 K century$^{-1}$. The elemental composition is obtained by multiplying the ice or gas abundance of each molecule with its elemental composition (e.g., for CH$_{3}$OH carbon = abundance $\times$ 1, hydrogen = abundance $\times$ 4, and oxygen = abundance $\times$ 1) and subsequently summing all contributions to an element. 

The resulting ice elemental compositions for [Hydrogen]/[Carbon], [Nitrogen]/[Carbon], [Oxygen]/[Carbon], and [Sulphur]/[Carbon] are shown in Fig. \ref{fig:elem_rat}. Figure \ref{fig:elem_rat_radius} in the Appendix shows the same plot, but with elemental ratios plotted versus protoplanetary disk radius by using an average protoplanetary disk temperature profile $T(r) = 200 \times (r / 1 AU)^{0.62}$ K \citep{andrews2007}. The elemental ratio of [Phosphorus]/[Carbon] is shown in Fig. \ref{fig:phosphor} in the Appendix. The bulk ice elemental ratios are plotted in blue, while the addition of a low (10$^{-4}\times$[H$_{2}$O]) and high (10$^{-3}\times$[H$_{2}$O]) fraction of organic molecules are plotted in red and green, respectively. The peak desorption temperatures of several bulk ice species are indicated, as well as the elemental composition of the extraterrestrial substances Soluble Organic Matter \citep[SOM,][]{sephton2002} and Insoluble Organic Matter \citep[IOM,][]{alexander2017}. Both SOM and IOM are regularly extracted from meteorites and SOM shows similarities with the organic material measured on several solar system objects, such as comet 67P/ Churyumov-Gerasimenko \citep{hanni2022}. In this work, we use the average elemental composition of C$_{100}$H$_{155}$N$_{3}$O$_{20}$S$_{3}$ for SOM obtained from the Murchison meteorite \citep{schmitt-kopplin2010} and the average IOM composition of C$_{100}$H$_{77}$O$_{14}$N$_{3}$S$_{2}$ presented in \citet{alexander2017}. 

First, we note that the addition of organic material affects the elemental ratios at low temperatures, when bulk ice dominates. The combined contribution of organic elemental carbon lowers all elemental ratios. The effect is most pronounced for the high fraction organic matter model where it can lower ratios by up to 100\%. In this scenario we only take the contribution of the 133 molecules studied in this work into account. In a natural environment the diversity of the organic components may be many times larger \citep[e.g.,][]{hanni2022} and therefore have an even larger impact on elemental ratios. 

Second, we note the importance of the organic material to set the ice elemental composition when the bulk ice species are removed from the solid state. In our model, all the bulk ice species are lost before $\sim$115~K, while the species that contribute to the nitrogen and sulphur budget are gone at lower temperatures of $\sim$100~K and $\sim$75~K, respectively. These differences in bulk ice desorption explain the dips seen in [N]/[C] and [S]/[C] between 75 and 100~K. At these points the nitrogen (NH$_{3}$) and sulphur (CS) reservoirs have desorbed, but the main carrier of carbon (CH$_{3}$OH) is still present. It is likely that these dips will be less pronounced in reality due to the trapping of the volatile nitrogen and sulphur carriers in the dominating water-phase, which desorbs together methanol. After desorption of the bulk ice reservoir the organic material starts to dominate the overall elemental composition on the dust grains.

The elemental ratios are biased towards the molecules included in this work and the assumption that molecules contribute equally to the total molecular budget is unlikely to be realistic. Therefore, care needs to be taken with any interpretation of Fig. \ref{fig:elem_rat}. However, some general observations can be made. First, at low temperature($\leq$115~K) or larger radii ($\geq$3~au), the bulk ices dominate the elemental budget and ratios are seen to vary in a stepwise manner. Objects formed in this regime will have predictable elemental compositions, depending on the ``step'' they form at. However, in the warmer disk regions closer to the (proto)star and where water has desorbed, the elemental budget is set by the large mixture of organic molecules, which desorb at different temperatures. Consequently, the elemental composition is seen to be much more varied, especially in [Hydrogen]/[Carbon] and [Oxygen]/[Carbon]. Objects that form in this regime could thus be expected to have a much more varied or less predictable elemental signature. 

It is interesting to point out that the organic molecules that remain on the grains at $T \geq$100~K show similarities with, and in some cases are identical to, the molecules used by \citet{kudo2002} to determine the sticking velocity of organics-coated grains. These authors found that organics-coated grains rapidly coagulate in the 2.3 -- 3.0 au region of a disk. This region is in good agreement with the location where we find most organic molecules still coating the grains after water has desorbed. 

Finally, the evolution of various functional chemical groups are plotted in Fig. \ref{fig:functional_groups}. The abundances of the functional groups have been determined in the same way that the elemental abundances have. Ratios of amides ([-NC(O)-]), cyanides ([-CN]), alcohols ([-OH]), carboxylic acids ([-COOH]), aldehydes + ketones ([-C(O)-]), esters ([-C(O)O-]), and ethers ([-O-]) with respect to amines ([-NH$_{2}$]) are plotted. This plot reflects several observations already made in Sect. \ref{sec:result} about the occurrence in of certain functional groups in the data set, for example that the alcohol (-OH) group is most abundant. It also shows that some of these functional groups can be quickly depleted as the environment warms up. In particular esters and ethers will be removed quickly. On the other hand, amides and amines seem to remain present to higher temperatures. In this way we see how chemical make-up and average properties of an organics-coated grain may change with temperature and could potentially be used as a tracers of thermal history of a particle in the solar system. However, it is important to acknowledge that this view is biased towards relatively small and volatile organic molecules. For example, IOM is rich in ether (-O-) cross links \citep{remusat2005b} and because this substance is very refractory it will ensure that the ether functional group will remain prominent component of the organic coating of grains at elevated temperatures. 

\section{Conclusion}

This study presents a large number of desorption parameters, that is desorption energies ($E_{\rm des}$) and pre-exponential frequency factors ($\nu$). These parameters will find use in astrochemical models and help to understand the evolution of the chemical and elemental composition in star and planet-forming regions. Because the list of molecules is dominated by organic molecules and salts of medium-volatility these data are of particular importance to assess the surface chemistry in regions were water ice has been removed from dust grains.

To expand the number of molecules for which desorption parameters are available, experimental temperature programmed desorption data have been collected from the literature and analysed with the Redhead Transition State Theory (Redhead-TST) method to determine the pre-exponential frequency factor ($\nu$) and the desorption energy ($E$). A comparison with literature $\nu$ and $E_{\rm des}$ values shows that the Redhead-TST method provides reliable results that are on par with results of rigorous experimental methods. We emphasise that the usage of accurately determined pre-factor values, instead of assumed values or the often used Hasagawa equation, is essential to properly simulate the desorption profiles of molecules. Due to the large amount of data collected in this study, trends can be searched for. No relationship between molecule mass or number of atoms is found, but a relationship between the pre-factor and molecule mass in the form of log$_{10}$($\nu$) = 2.65ln($m$) + 8.07 is, which can be used to determine this parameter in future studies. Mean desorption parameters are provided and used to highlight how the desorption of these species can affect chemical and elemental compositions. 

\begin{acknowledgements}
The authors thank E.F. van Dishoeck, E.G. B{\o}gelund, and C. Ceccarelli for helpful discussions and feedback. Thanks go out to K.-J. Chuang for making unpublished TPD plots of CH$_{2}$CHNH$_{2}$ and CH$_{3}$CHNH available for analysis. The authors thank the many astrochemists and surface scientists who have contributed data to the literature on which this study relies. N.F.W.L. acknowledges support from the Swiss National Science Foundation (SNSF) Ambizione grant 193453 and NCCR PlanetS. M.M. acknowledges the French national programme “Physique et Chimie du Milieu Interstellaire” (PCMI) of CNRS/INSU with INC/INP cofunded by CEA and CNES.
\end{acknowledgements}

\bibliographystyle{aa} 
\bibliography{bib} 

\begin{thebibliography}{203}
\expandafter\ifx\csname natexlab\endcsname\relax\def\natexlab#1{#1}\fi

\bibitem[{Abplanalp {et~al.}(2015)Abplanalp, Borsuk, Jones, \&
  Kaiser}]{Abplanalp2015}
Abplanalp, M.~J., Borsuk, A., Jones, B.~M., \& Kaiser, R.~I. 2015, ApJ, 814, 45

\bibitem[{Abplanalp {et~al.}(2016)Abplanalp, F{\"o}rstel, \&
  Kaiser}]{abplanalp2016b}
Abplanalp, M.~J., F{\"o}rstel, M., \& Kaiser, R.~I. 2016, Chem. Phys. Lett.,
  644, 79

\bibitem[{Abplanalp {et~al.}(2019)Abplanalp, Frigge, \&
  Kaiser}]{Abplanalp2019a}
Abplanalp, M.~J., Frigge, R., \& Kaiser, R.~I. 2019, Sci. Adv., 5, eaaw5841

\bibitem[{Abplanalp \& Kaiser(2019)}]{abplanalp2019c}
Abplanalp, M.~J. \& Kaiser, R.~I. 2019, PCCP, 21, 16949

\bibitem[{Alexander {et~al.}(2017)Alexander, Cody, De~Gregorio, Nittler, \&
  Stroud}]{alexander2017}
Alexander, C.~O., Cody, G., De~Gregorio, B., Nittler, L., \& Stroud, R. 2017,
  Geochemistry, 77, 227

\bibitem[{Altwegg {et~al.}(2016)Altwegg, Balsiger, Bar-Nun, Berthelier, Bieler,
  Bochsler, Briois, Calmonte, Combi, Cottin, {et~al.}}]{altwegg2016}
Altwegg, K., Balsiger, H., Bar-Nun, A., {et~al.} 2016, Sci. Adv., 2, e1600285

\bibitem[{Andrews \& Williams(2007)}]{andrews2007}
Andrews, S.~M. \& Williams, J.~P. 2007, ApJ, 659, 705

\bibitem[{Bahr \& Kempter(2007)}]{Bahr2007a}
Bahr, S. \& Kempter, V. 2007, J. Chem. Phys., 127, 074707

\bibitem[{Behmard {et~al.}(2019)Behmard, Fayolle, Graninger, Bergner,
  Mart{\'\i}n-Dom{\'e}nech, Maksyutenko, Rajappan, \& {\"O}berg}]{behmard2019}
Behmard, A., Fayolle, E.~C., Graninger, D.~M., {et~al.} 2019, ApJ, 875, 73

\bibitem[{Bergantini {et~al.}(2018{\natexlab{a}})Bergantini, Abplanalp,
  Pokhilko, Krylov, Shingledecker, Herbst, \& Kaiser}]{Bergantini2018b}
Bergantini, A., Abplanalp, M.~J., Pokhilko, P., {et~al.} 2018{\natexlab{a}},
  ApJ, 860, 108

\bibitem[{Bergantini {et~al.}(2018{\natexlab{b}})Bergantini, Frigge, \&
  Kaiser}]{Bergantini2018d}
Bergantini, A., Frigge, R., \& Kaiser, R.~I. 2018{\natexlab{b}}, ApJ, 859, 59

\bibitem[{Bergantini {et~al.}(2018{\natexlab{c}})Bergantini, G{\'o}bi,
  Abplanalp, \& Kaiser}]{Bergantini2018c}
Bergantini, A., G{\'o}bi, S., Abplanalp, M.~J., \& Kaiser, R.~I.
  2018{\natexlab{c}}, ApJ, 852, 70

\bibitem[{Bergantini {et~al.}(2018{\natexlab{d}})Bergantini, Zhu, \&
  Kaiser}]{Bergantini2018a}
Bergantini, A., Zhu, C., \& Kaiser, R.~I. 2018{\natexlab{d}}, ApJ, 862, 140

\bibitem[{Bergner {et~al.}(2016)Bergner, {\"O}berg, Rajappan, \&
  Fayolle}]{Bergner2016}
Bergner, J.~B., {\"O}berg, K.~I., Rajappan, M., \& Fayolle, E.~C. 2016, ApJ,
  829, 85

\bibitem[{Bertin {et~al.}(2017)Bertin, Doronin, Fillion, Michaut, Philippe,
  Lattelais, Markovits, Pauzat, Ellinger, \& Guillemin}]{Bertin2017}
Bertin, M., Doronin, M., Fillion, J.-H., {et~al.} 2017, A\&A, 598, A18

\bibitem[{Bertin {et~al.}(2011)Bertin, Romanzin, Michaut, Jeseck, \&
  Fillion}]{bertin2011}
Bertin, M., Romanzin, C., Michaut, X., Jeseck, P., \& Fillion, J.-H. 2011, J.
  Phys. Chem. C, 115, 12920

\bibitem[{Bertrand {et~al.}(2022)Bertrand, Lellouch, Holler, Young, Schmitt,
  Oliveira, Sicardy, Forget, Grundy, Merlin, {et~al.}}]{bertrand2022}
Bertrand, T., Lellouch, E., Holler, B., {et~al.} 2022, Icarus, 373, 114764

\bibitem[{Bianchi {et~al.}(2022)Bianchi, L{\'o}pez-Sepulcre, Ceccarelli,
  Codella, Podio, Bouvier, \& Enrique-Romero}]{bianchi2022}
Bianchi, E., L{\'o}pez-Sepulcre, A., Ceccarelli, C., {et~al.} 2022, The
  Astrophysical Journal Letters, 928, L3

\bibitem[{Bisschop {et~al.}(2007)Bisschop, Fuchs, Boogert, Van~Dishoeck, \&
  Linnartz}]{Bisschop2007}
Bisschop, S., Fuchs, G., Boogert, A., Van~Dishoeck, E., \& Linnartz, H. 2007,
  A\&A, 470, 749

\bibitem[{Boamah {et~al.}(2014)Boamah, Sullivan, Shulenberger, Soe, Jacob,
  Yhee, Atkinson, Boyer, Haines, \& Arumainayagam}]{Boamah2014}
Boamah, M.~D., Sullivan, K.~K., Shulenberger, K.~E., {et~al.} 2014, Faraday
  Discuss., 168, 249

\bibitem[{B{\o}gelund {et~al.}(2019)B{\o}gelund, Barr, Taquet, Ligterink,
  Persson, Hogerheijde, \& van Dishoeck}]{bogelund2019}
B{\o}gelund, E.~G., Barr, A.~G., Taquet, V., {et~al.} 2019, A\&A, 628, A2

\bibitem[{Boogert {et~al.}(2015)Boogert, Gerakines, \& Whittet}]{boogert2015}
Boogert, A.~A., Gerakines, P.~A., \& Whittet, D.~C. 2015, ARA\&A, 53, 541

\bibitem[{Boogert {et~al.}(2008)Boogert, Pontoppidan, Knez, Lahuis,
  Kessler-Silacci, van Dishoeck, Blake, Augereau, Bisschop, Bottinelli,
  {et~al.}}]{boogert2008}
Boogert, A.~C., Pontoppidan, K.~M., Knez, C., {et~al.} 2008, The Astrophysical
  Journal, 678, 985

\bibitem[{Bosman {et~al.}(2019)Bosman, Cridland, \& Miguel}]{bosman2019}
Bosman, A.~D., Cridland, A.~J., \& Miguel, Y. 2019, A\&A, 632, L11

\bibitem[{Bossa {et~al.}(2009{\natexlab{a}})Bossa, Theule, Duvernay, \&
  Chiavassa}]{Bossa2009b}
Bossa, J., Theule, P., Duvernay, F., \& Chiavassa, T. 2009{\natexlab{a}}, ApJ,
  707, 1524

\bibitem[{Bossa {et~al.}(2012)Bossa, Borget, Duvernay, Danger, Theul{\'e}, \&
  Chiavassa}]{Bossa2012}
Bossa, J.-B., Borget, F., Duvernay, F., {et~al.} 2012, Aust. J. Chem., 65, 129

\bibitem[{Bossa {et~al.}(2009{\natexlab{b}})Bossa, Duvernay, Theul{\'e},
  Borget, d'Hendecourt, \& Chiavassa}]{bossa2009}
Bossa, J.-B., Duvernay, F., Theul{\'e}, P., {et~al.} 2009{\natexlab{b}}, A\&A,
  506, 601

\bibitem[{Bovolenta {et~al.}(2022)Bovolenta, Vogt-Geisse, Bovino, \&
  Grassi}]{bovolenta2022}
Bovolenta, G.~M., Vogt-Geisse, S., Bovino, S., \& Grassi, T. 2022, ApJS, 262,
  17

\bibitem[{Burke {et~al.}(2008)Burke, Wolff, Edridge, \& Brown}]{Burke2008}
Burke, D., Wolff, A., Edridge, J., \& Brown, W. 2008, J. Chem. Phys., 128,
  104702

\bibitem[{Burke \& Brown(2010)}]{burke2010}
Burke, D.~J. \& Brown, W.~A. 2010, PCCP, 12, 5947

\bibitem[{Burke {et~al.}(2015{\natexlab{a}})Burke, Puletti, Brown, Woods, Viti,
  \& Slater}]{burke2015a}
Burke, D.~J., Puletti, F., Brown, W.~A., {et~al.} 2015{\natexlab{a}}, MNRAS,
  447, 1444

\bibitem[{Burke {et~al.}(2015{\natexlab{b}})Burke, Puletti, Woods, Viti,
  Slater, \& Brown}]{burke2015b}
Burke, D.~J., Puletti, F., Woods, P.~M., {et~al.} 2015{\natexlab{b}}, J. Phys.
  Chem. A, 119, 6837

\bibitem[{Burke {et~al.}(2015{\natexlab{c}})Burke, Puletti, Woods, Viti,
  Slater, \& Brown}]{burke2015c}
Burke, D.~J., Puletti, F., Woods, P.~M., {et~al.} 2015{\natexlab{c}}, J. Chem.
  Phys., 143, 164704

\bibitem[{Butscher {et~al.}(2016)Butscher, Duvernay, Danger, \&
  Chiavassa}]{Butscher2016}
Butscher, T., Duvernay, F., Danger, G., \& Chiavassa, T. 2016, A\&A, 593, A60

\bibitem[{Carrascosa {et~al.}(2020)Carrascosa, Cruz-D{\'\i}az, Mu{\~n}oz~Caro,
  Dartois, \& Chen}]{Carrascosa2020}
Carrascosa, H., Cruz-D{\'\i}az, G., Mu{\~n}oz~Caro, G.~M., Dartois, E., \&
  Chen, Y. 2020, MNRAS, 493, 821

\bibitem[{Ceccarelli {et~al.}(2022)Ceccarelli, Codella, Balucani,
  Bockel{\'e}e-Morvan, Herbst, Vastel, Caselli, Favre, Lefloch, \&
  {\"O}berg}]{ceccarelli2022}
Ceccarelli, C., Codella, C., Balucani, N., {et~al.} 2022, arXiv preprint
  arXiv:2206.13270

\bibitem[{Cernicharo {et~al.}(2022)Cernicharo, Fuentetaja, Cabezas,
  Ag{\'u}ndez, Marcelino, Tercero, Pardo, \& de~Vicente}]{cernicharo2022}
Cernicharo, J., Fuentetaja, R., Cabezas, C., {et~al.} 2022, A\&A, 663, L5

\bibitem[{Cernicharo {et~al.}(2020)Cernicharo, Marcelino, Ag{\'u}ndez,
  Berm{\'u}dez, Cabezas, Tercero, \& Pardo}]{cernicharo2020}
Cernicharo, J., Marcelino, N., Ag{\'u}ndez, M., {et~al.} 2020, A\&A, 642, L8

\bibitem[{Chaabouni {et~al.}(2020)Chaabouni, Baouche, Diana, \&
  Minissale}]{Chaabouni2020}
Chaabouni, H., Baouche, S., Diana, S., \& Minissale, M. 2020, A\&A, 636, A4

\bibitem[{Chaabouni {et~al.}(2018)Chaabouni, Diana, Nguyen, \&
  Dulieu}]{chaabouni2018}
Chaabouni, H., Diana, S., Nguyen, T., \& Dulieu, F. 2018, A\&A, 612, A47

\bibitem[{Chuang {et~al.}(2020)Chuang, Fedoseev, Qasim, Ioppolo, J{\"a}ger,
  Henning, Palumbo, van Dishoeck, \& Linnartz}]{Chuang2020}
Chuang, K.-J., Fedoseev, G., Qasim, D., {et~al.} 2020, A\&A, 635, A199

\bibitem[{Congiu {et~al.}(2012)Congiu, Fedoseev, Ioppolo, Dulieu, Chaabouni,
  Baouche, Lemaire, Laffon, Parent, Lamberts, {et~al.}}]{Congiu2012a}
Congiu, E., Fedoseev, G., Ioppolo, S., {et~al.} 2012, ApJL, 750, L12

\bibitem[{Corazzi {et~al.}(2021)Corazzi, Brucato, Poggiali, Podio, Fedele, \&
  Codella}]{corazzi2021}
Corazzi, M.~A., Brucato, J.~R., Poggiali, G., {et~al.} 2021, ApJ, 913, 128

\bibitem[{Coupeaud {et~al.}(2008)Coupeaud, Pi{\'e}tri, Allouche, Aycard, \&
  Couturier-Tamburelli}]{Coupeaud2008}
Coupeaud, A., Pi{\'e}tri, N., Allouche, A., Aycard, J.-P., \&
  Couturier-Tamburelli, I. 2008, J. Phys. Chem. A, 112, 8024

\bibitem[{Couturier-Tamburelli {et~al.}(2018)Couturier-Tamburelli, Toumi,
  Pi{\'e}tri, \& Chiavassa}]{couturier-tamburelli2018}
Couturier-Tamburelli, I., Toumi, A., Pi{\'e}tri, N., \& Chiavassa, T. 2018,
  Icarus, 300, 477

\bibitem[{Danger {et~al.}(2011{\natexlab{a}})Danger, Borget, Chomat, Duvernay,
  Theul{\'e}, Guillemin, d’Hendecourt, \& Chiavassa}]{Danger2011a}
Danger, G., Borget, F., Chomat, M., {et~al.} 2011{\natexlab{a}}, A\&A, 535, A47

\bibitem[{Danger {et~al.}(2011{\natexlab{b}})Danger, Bossa, De~Marcellus,
  Borget, Duvernay, Theul{\'e}, Chiavassa, \& d’Hendecourt}]{Danger2011b}
Danger, G., Bossa, J.-B., De~Marcellus, P., {et~al.} 2011{\natexlab{b}}, A\&A,
  525, A30

\bibitem[{Danger {et~al.}(2012)Danger, Duvernay, Theul{\'e}, Borget, \&
  Chiavassa}]{Danger2012}
Danger, G., Duvernay, F., Theul{\'e}, P., Borget, F., \& Chiavassa, T. 2012,
  ApJ, 756, 11

\bibitem[{Danger {et~al.}(2014)Danger, Rimola, Abou~Mrad, Duvernay, Roussin,
  Theule, \& Chiavassa}]{Danger2014}
Danger, G., Rimola, A., Abou~Mrad, N., {et~al.} 2014, PCCP, 16, 3360

\bibitem[{De~Jong \& Niemantsverdriet(1990{\natexlab{a}})}]{dejong1990b}
De~Jong, A. \& Niemantsverdriet, J. 1990{\natexlab{a}}, Vacuum, 41, 232

\bibitem[{De~Jong \& Niemantsverdriet(1990{\natexlab{b}})}]{dejong1990a}
De~Jong, A. \& Niemantsverdriet, J. 1990{\natexlab{b}}, Surf. Sci., 233, 355

\bibitem[{Demers {et~al.}(2002)Demers, {\"O}stblom, Zhang, Jang, Liedberg, \&
  Mirkin}]{Demers2002}
Demers, L.~M., {\"O}stblom, M., Zhang, H., {et~al.} 2002, J. Am. Chem. Soc.,
  124, 11248

\bibitem[{DeSimone {et~al.}(2013)DeSimone, Olanrewaju, Grieves, \&
  Orlando}]{DeSimone2013}
DeSimone, A.~J., Olanrewaju, B.~O., Grieves, G.~A., \& Orlando, T.~M. 2013, J.
  Chem. Phys., 138, 084703

\bibitem[{Dostert {et~al.}(2016)Dostert, O'Brien, Mirabella, Ivars-Barcel{\'o},
  \& Schauermann}]{Dostert2016}
Dostert, K.-H., O'Brien, C.~P., Mirabella, F., Ivars-Barcel{\'o}, F., \&
  Schauermann, S. 2016, PCCP, 18, 13960

\bibitem[{Duvernay {et~al.}(2014)Duvernay, Danger, Theul{\'e}, Chiavassa, \&
  Rimola}]{Duvernay2014}
Duvernay, F., Danger, G., Theul{\'e}, P., Chiavassa, T., \& Rimola, A. 2014,
  ApJ, 791, 75

\bibitem[{Duvernay {et~al.}(2010)Duvernay, Dufauret, Danger, Theul{\'e},
  Borget, \& Chiavassa}]{Duvernay2010}
Duvernay, F., Dufauret, V., Danger, G., {et~al.} 2010, A\&A, 523, A79

\bibitem[{Eckhardt {et~al.}(2019)Eckhardt, Bergantini, Singh, Schreiner, \&
  Kaiser}]{Eckhardt2019}
Eckhardt, A.~K., Bergantini, A., Singh, S.~K., Schreiner, P.~R., \& Kaiser,
  R.~I. 2019, Angew. Chem., 131, 5719

\bibitem[{Fedoseev {et~al.}(2017)Fedoseev, Chuang, Ioppolo, Qasim, van
  Dishoeck, \& Linnartz}]{Fedoseev2017}
Fedoseev, G., Chuang, K.-J., Ioppolo, S., {et~al.} 2017, ApJ, 842, 52

\bibitem[{Fedoseev {et~al.}(2016)Fedoseev, Chuang, van Dishoeck, Ioppolo, \&
  Linnartz}]{Fedoseev2016}
Fedoseev, G., Chuang, K.-J., van Dishoeck, E.~F., Ioppolo, S., \& Linnartz, H.
  2016, MNRAS, 460, 4297

\bibitem[{Fedoseev {et~al.}(2015{\natexlab{a}})Fedoseev, Cuppen, Ioppolo,
  Lamberts, \& Linnartz}]{Fedoseev2015a}
Fedoseev, G., Cuppen, H.~M., Ioppolo, S., Lamberts, T., \& Linnartz, H.
  2015{\natexlab{a}}, MNRAS, 448, 1288

\bibitem[{Fedoseev {et~al.}(2015{\natexlab{b}})Fedoseev, Ioppolo, Zhao,
  Lamberts, \& Linnartz}]{Fedoseev2015b}
Fedoseev, G., Ioppolo, S., Zhao, D., Lamberts, T., \& Linnartz, H.
  2015{\natexlab{b}}, MNRAS, 446, 439

\bibitem[{Ferrero {et~al.}(2022)Ferrero, Grieco, Ibrahim~Mohamed, Dulieu,
  Rimola, Ceccarelli, Nervi, Minissale, \& Ugliengo}]{ferrero2022}
Ferrero, S., Grieco, F., Ibrahim~Mohamed, A., {et~al.} 2022, MNRAS, 516, 2586

\bibitem[{F{\"o}rstel {et~al.}(2017)F{\"o}rstel, Bergantini, Maksyutenko,
  G{\'o}bi, \& Kaiser}]{Forstel2017}
F{\"o}rstel, M., Bergantini, A., Maksyutenko, P., G{\'o}bi, S., \& Kaiser,
  R.~I. 2017, ApJ, 845, 83

\bibitem[{F{\"o}rstel {et~al.}(2016{\natexlab{a}})F{\"o}rstel, Maksyutenko,
  Jones, Sun, Chang, \& Kaiser}]{Forstel2016a}
F{\"o}rstel, M., Maksyutenko, P., Jones, B., {et~al.} 2016{\natexlab{a}},
  Chemical Commun., 52, 741

\bibitem[{F{\"o}rstel {et~al.}(2016{\natexlab{b}})F{\"o}rstel, Tsegaw,
  Maksyutenko, Mebel, Sander, \& Kaiser}]{Forstel2016b}
F{\"o}rstel, M., Tsegaw, Y.~A., Maksyutenko, P., {et~al.} 2016{\natexlab{b}},
  ChemPhysChem, 17, 2726

\bibitem[{Fray \& Schmitt(2009)}]{fray2009}
Fray, N. \& Schmitt, B. 2009, Planetary and Space Science, 57, 2053

\bibitem[{Frigge {et~al.}(2018{\natexlab{a}})Frigge, Zhu, Turner, Abplanalp,
  Bergantini, Sun, Chen, Chang, \& Kaiser}]{Frigge2018b}
Frigge, R., Zhu, C., Turner, A.~M., {et~al.} 2018{\natexlab{a}}, ApJ, 862, 84

\bibitem[{Frigge {et~al.}(2018{\natexlab{b}})Frigge, Zhu, Turner, Abplanalp,
  Sun, Huang, Chang, \& Kaiser}]{Frigge2018a}
Frigge, R., Zhu, C., Turner, A.~M., {et~al.} 2018{\natexlab{b}}, Chemical
  Commun., 54, 10152

\bibitem[{Garrod {et~al.}(2022)Garrod, Jin, Matis, Jones, Willis, \&
  Herbst}]{garrod2022}
Garrod, R.~T., Jin, M., Matis, K.~A., {et~al.} 2022, The Astrophysical Journal
  Supplement Series, 259, 1

\bibitem[{Germain {et~al.}(2022)Germain, Tinacci, Pantaleone, Ceccarelli, \&
  Ugliengo}]{germain2022}
Germain, A., Tinacci, L., Pantaleone, S., Ceccarelli, C., \& Ugliengo, P. 2022,
  ACS Earth Space Chem., 6, 1286

\bibitem[{Gibb \& Whittet(2002)}]{gibb2002}
Gibb, E. \& Whittet, D. 2002, The Astrophysical Journal, 566, L113

\bibitem[{Goesmann {et~al.}(2015)Goesmann, Rosenbauer, Bredeh{\"o}ft, Cabane,
  Ehrenfreund, Gautier, Giri, Kr{\"u}ger, Le~Roy, MacDermott,
  {et~al.}}]{goesmann2015}
Goesmann, F., Rosenbauer, H., Bredeh{\"o}ft, J.~H., {et~al.} 2015, Sci., 349,
  aab0689

\bibitem[{Gorai {et~al.}(2020)Gorai, Bhat, Sil, Mondal, Ghosh, Chakrabarti, \&
  Das}]{gorai2020}
Gorai, P., Bhat, B., Sil, M., {et~al.} 2020, ApJ, 895, 86

\bibitem[{Guennoun {et~al.}(2005)Guennoun, Couturier-Tamburelli, Pi{\'e}tri, \&
  Aycard}]{Guennoun2005a}
Guennoun, Z., Couturier-Tamburelli, I., Pi{\'e}tri, N., \& Aycard, J.-P. 2005,
  J. Phys. Chem. B, 109, 3437

\bibitem[{H{\"a}nni {et~al.}(2022)H{\"a}nni, Altwegg, Combi, Fuselier,
  De~Keyser, Rubin, \& Wampfler}]{hanni2022}
H{\"a}nni, N., Altwegg, K., Combi, M., {et~al.} 2022, Nat. Commun., 13, 1

\bibitem[{Hasegawa {et~al.}(1992)Hasegawa, Herbst, \& Leung}]{hasegawa1992}
Hasegawa, T.~I., Herbst, E., \& Leung, C.~M. 1992, ApJS, 82, 167

\bibitem[{He {et~al.}(2016)He, Acharyya, \& Vidali}]{he2016b}
He, J., Acharyya, K., \& Vidali, G. 2016, ApJ, 825, 89

\bibitem[{He {et~al.}(2015)He, Vidali, Lemaire, \& Garrod}]{He2015b}
He, J., Vidali, G., Lemaire, J.-L., \& Garrod, R.~T. 2015, ApJ, 799, 49

\bibitem[{Heyl {et~al.}(2022)Heyl, Holdship, \& Viti}]{heyl2022}
Heyl, J., Holdship, J., \& Viti, S. 2022, ApJ, 931, 26

\bibitem[{Hsu {et~al.}(2022)Hsu, Liu, Liu, Sahu, Lee, Tatematsu, Kim, Hirano,
  Yang, Johnstone, {et~al.}}]{hsu2022}
Hsu, S.-Y., Liu, S.-Y., Liu, T., {et~al.} 2022, The Astrophysical Journal, 927,
  218

\bibitem[{Ioppolo {et~al.}(2021)Ioppolo, Fedoseev, Chuang, Cuppen, Clements,
  Jin, Garrod, Qasim, Kofman, van Dishoeck, {et~al.}}]{Ioppolo2021}
Ioppolo, S., Fedoseev, G., Chuang, K.-J., {et~al.} 2021, Nature Astronomy, 5,
  197

\bibitem[{Ioppolo {et~al.}(2014)Ioppolo, Fedoseev, Minissale, Congiu, Dulieu,
  \& Linnartz}]{ioppolo2014}
Ioppolo, S., Fedoseev, G., Minissale, M., {et~al.} 2014, PCCP, 16, 8270

\bibitem[{Ioppolo {et~al.}(2011)Ioppolo, Van~Boheemen, Cuppen, Van~Dishoeck, \&
  Linnartz}]{Ioppolo2011}
Ioppolo, S., Van~Boheemen, Y., Cuppen, H., Van~Dishoeck, E., \& Linnartz, H.
  2011, MNRAS, 413, 2281

\bibitem[{Jimenez-Escobar \& Mu{\~n}oz~Caro(2011)}]{jimenez-escobar2011}
Jimenez-Escobar, A. \& Mu{\~n}oz~Caro, G. 2011, The Molecular Universe, 280,
  153

\bibitem[{Jin \& Garrod(2020)}]{jin2020}
Jin, M. \& Garrod, R.~T. 2020, The Astrophysical Journal Supplement Series,
  249, 26

\bibitem[{Johnson {et~al.}(2021)Johnson, Young, Protopapa, Schmitt, Gabasova,
  Lewis, Stansberry, Mandt, \& White}]{johnson2021}
Johnson, P.~E., Young, L.~A., Protopapa, S., {et~al.} 2021, Icarus, 356, 114070

\bibitem[{Jones {et~al.}(2014)Jones, Kaiser, \& Strazzulla}]{Jones2014}
Jones, B.~M., Kaiser, R.~I., \& Strazzulla, G. 2014, ApJ, 788, 170

\bibitem[{Kaiser \& Maksyutenko(2015)}]{Kaiser2015a}
Kaiser, R.~I. \& Maksyutenko, P. 2015, Chem. Phys. Lett., 631, 59

\bibitem[{Keane {et~al.}(2001)Keane, Tielens, Boogert, Schutte, \&
  Whittet}]{keane2001b}
Keane, J., Tielens, A., Boogert, A., Schutte, W., \& Whittet, D. 2001, A\&A,
  376, 254

\bibitem[{Kimber {et~al.}(2018)Kimber, Toscano, \& Price}]{kimber2018}
Kimber, H.~J., Toscano, J., \& Price, S.~D. 2018, MNRAS, 476, 5332

\bibitem[{King(1975)}]{king1975}
King, D.~A. 1975, Surf. Sci., 47, 384

\bibitem[{Kleimeier {et~al.}(2021{\natexlab{a}})Kleimeier, Abplanalp, Johnson,
  Gozem, Wandishin, Shingledecker, \& Kaiser}]{kleimeier2021b}
Kleimeier, N.~F., Abplanalp, M.~J., Johnson, R.~N., {et~al.}
  2021{\natexlab{a}}, ApJ, 911, 24

\bibitem[{Kleimeier {et~al.}(2021{\natexlab{b}})Kleimeier, Eckhardt, \&
  Kaiser}]{Kleimeier2021a}
Kleimeier, N.~F., Eckhardt, A.~K., \& Kaiser, R.~I. 2021{\natexlab{b}}, J. Am.
  Chem. Soc., 143, 14009

\bibitem[{Kleimeier {et~al.}(2020{\natexlab{a}})Kleimeier, Eckhardt, Schreiner,
  \& Kaiser}]{Kleimeier2020b}
Kleimeier, N.~F., Eckhardt, A.~K., Schreiner, P.~R., \& Kaiser, R.~I.
  2020{\natexlab{a}}, Chem, 6, 3385

\bibitem[{Kleimeier \& Kaiser(2021)}]{kleimeier2021d}
Kleimeier, N.~F. \& Kaiser, R.~I. 2021, J. Phys. Chem. Lett., 13, 229

\bibitem[{Kleimeier {et~al.}(2020{\natexlab{b}})Kleimeier, Turner, Fortenberry,
  \& Kaiser}]{Kleimeier2020c}
Kleimeier, N.~F., Turner, A.~M., Fortenberry, R.~C., \& Kaiser, R.~I.
  2020{\natexlab{b}}, ChemPhysChem, 21, 1531

\bibitem[{Kruczkiewicz {et~al.}(2021)Kruczkiewicz, Vitorino, Congiu,
  Theul{\'e}, \& Dulieu}]{kruczkiewicz2021}
Kruczkiewicz, F., Vitorino, J., Congiu, E., Theul{\'e}, P., \& Dulieu, F. 2021,
  A\&A, 652, A29

\bibitem[{Kudo {et~al.}(2002)Kudo, Kouchi, Arakawa, \& Nakano}]{kudo2002}
Kudo, T., Kouchi, A., Arakawa, M., \& Nakano, H. 2002, Meteoritics \& Planetary
  Science, 37, 1975

\bibitem[{Kulterer {et~al.}(2020)Kulterer, Drozdovskaya, Coutens, Manigand, \&
  St{\'e}phan}]{kulterer2020}
Kulterer, B.~M., Drozdovskaya, M.~N., Coutens, A., Manigand, S., \&
  St{\'e}phan, G. 2020, MNRAS, 498, 276

\bibitem[{Lacy {et~al.}(1984)Lacy, Baas, Allamandola, Persson, McGregor,
  Lonsdale, Geballe, \& Van~de Bult}]{lacy1984}
Lacy, J., Baas, F., Allamandola, L., {et~al.} 1984, ApJ, 276, 533

\bibitem[{Lasne {et~al.}(2012)Lasne, Laffon, \& Parent}]{lasne2012}
Lasne, J., Laffon, C., \& Parent, P. 2012, PCCP, 14, 697

\bibitem[{Lattelais {et~al.}(2011)Lattelais, Bertin, Mokrane, Romanzin,
  Michaut, Jeseck, Fillion, Chaabouni, Congiu, Dulieu,
  {et~al.}}]{lattelais2011}
Lattelais, M., Bertin, M., Mokrane, H., {et~al.} 2011, A\&A, 532, A12

\bibitem[{Layssac {et~al.}(2020)Layssac, Guti{\'e}rrez-Quintanilla, Chiavassa,
  \& Duvernay}]{Layssac2020}
Layssac, Y., Guti{\'e}rrez-Quintanilla, A., Chiavassa, T., \& Duvernay, F.
  2020, MNRAS, 496, 5292

\bibitem[{Lee {et~al.}(2022)Lee, Codella, Ceccarelli, \&
  L{\'o}pez-Sepulcre}]{lee2022}
Lee, C.-F., Codella, C., Ceccarelli, C., \& L{\'o}pez-Sepulcre, A. 2022, ApJ,
  937, 10

\bibitem[{Lee {et~al.}(2021{\natexlab{a}})Lee, Changala, Loomis, Burkhardt,
  Xue, Cordiner, Charnley, McCarthy, \& McGuire}]{lee2021b}
Lee, K. L.~K., Changala, P.~B., Loomis, R.~A., {et~al.} 2021{\natexlab{a}},
  ApJL, 910, L2

\bibitem[{Lee {et~al.}(2021{\natexlab{b}})Lee, Loomis, Burkhardt, Cooke, Xue,
  Siebert, Shingledecker, Remijan, Charnley, McCarthy, {et~al.}}]{lee2021a}
Lee, K. L.~K., Loomis, R.~A., Burkhardt, A.~M., {et~al.} 2021{\natexlab{b}},
  ApJL, 908, L11

\bibitem[{Leroux {et~al.}(2021)Leroux, Guillemin, \& Krim}]{Leroux2021c}
Leroux, K., Guillemin, J.-C., \& Krim, L. 2021, MNRAS, 507, 2632

\bibitem[{Li {et~al.}(2021)Li, Bergin, Blake, Ciesla, \&
  Hirschmann}]{li2021carbon}
Li, J., Bergin, E.~A., Blake, G.~A., Ciesla, F.~J., \& Hirschmann, M.~M. 2021,
  Sci. Adv., 7, eabd3632

\bibitem[{Ligterink {et~al.}(2021)Ligterink, Ahmadi, Coutens, Calcutt, van
  Dishoeck, Linnartz, J{\o}rgensen, Garrod, Bouwman, {et~al.}}]{ligterink2021}
Ligterink, N., Ahmadi, A., Coutens, A., {et~al.} 2021, A\&A, 647, A87

\bibitem[{Ligterink {et~al.}(2017)Ligterink, Coutens, Kofman, M{\"u}ller,
  Garrod, Calcutt, Wampfler, J{\o}rgensen, Linnartz, \&
  Van~Dishoeck}]{ligterink2017}
Ligterink, N., Coutens, A., Kofman, V., {et~al.} 2017, MNRAS, 469, 2219

\bibitem[{Ligterink {et~al.}(2018{\natexlab{a}})Ligterink, Terwisscha~van
  Scheltinga, Taquet, J{\o}rgensen, Cazaux, van Dishoeck, \&
  Linnartz}]{ligterink2018a}
Ligterink, N., Terwisscha~van Scheltinga, J., Taquet, V., {et~al.}
  2018{\natexlab{a}}, MNRAS, 480, 3628

\bibitem[{Ligterink {et~al.}(2022)Ligterink, Ahmadi, Luitel, Coutens, Calcutt,
  Tychoniec, Linnartz, J{\o}rgensen, Garrod, \& Bouwman}]{ligterink2022}
Ligterink, N.~F., Ahmadi, A., Luitel, B., {et~al.} 2022, ACS Earth Space Chem.,
  6, 455

\bibitem[{Ligterink {et~al.}(2020)Ligterink, El-Abd, Brogan, Hunter, Remijan,
  Garrod, \& McGuire}]{ligterink2020b}
Ligterink, N.~F., El-Abd, S.~J., Brogan, C.~L., {et~al.} 2020, ApJ, 901, 37

\bibitem[{Ligterink {et~al.}(2018{\natexlab{b}})Ligterink, Calcutt, Coutens,
  Kristensen, Bourke, Drozdovskaya, M{\"u}ller, Wampfler, van~der Wiel,
  Van~Dishoeck, {et~al.}}]{ligterink2018b}
Ligterink, N. F.~W., Calcutt, H., Coutens, A., {et~al.} 2018{\natexlab{b}},
  A\&A, 619, A28

\bibitem[{Liu {et~al.}(2002)Liu, Rodriguez, Chang, Hrbek, \&
  Gonz{\'a}lez}]{Liu2002b}
Liu, G., Rodriguez, J., Chang, Z., Hrbek, J., \& Gonz{\'a}lez, L. 2002, J.
  Phys. Chem. B, 106, 9883

\bibitem[{Loomis {et~al.}(2021)Loomis, Burkhardt, Shingledecker, Charnley,
  Cordiner, Herbst, Kalenskii, Lee, Willis, Xue, {et~al.}}]{loomis2021}
Loomis, R.~A., Burkhardt, A.~M., Shingledecker, C.~N., {et~al.} 2021, Nature
  Astronomy, 5, 188

\bibitem[{Mahjoub {et~al.}(2017)Mahjoub, Poston, Blacksberg, Eiler, Brown,
  Ehlmann, Hodyss, Hand, Carlson, \& Choukroun}]{Mahjoub2017}
Mahjoub, A., Poston, M.~J., Blacksberg, J., {et~al.} 2017, ApJ, 846, 148

\bibitem[{Maity {et~al.}(2014{\natexlab{a}})Maity, Kaiser, \&
  Jones}]{Maity2014b}
Maity, S., Kaiser, R.~I., \& Jones, B.~M. 2014{\natexlab{a}}, Faraday Discuss.,
  168, 485

\bibitem[{Maity {et~al.}(2014{\natexlab{b}})Maity, Kaiser, \&
  Jones}]{Maity2014a}
Maity, S., Kaiser, R.~I., \& Jones, B.~M. 2014{\natexlab{b}}, Faraday Discuss.,
  168, 485

\bibitem[{Maity {et~al.}(2015)Maity, Kaiser, \& Jones}]{Maity2015}
Maity, S., Kaiser, R.~I., \& Jones, B.~M. 2015, PCCP, 17, 3081

\bibitem[{Maksyutenko {et~al.}(2022)Maksyutenko, Mart{\'\i}n-Dom{\'e}nech,
  Piacentino, {\"O}berg, \& Rajappan}]{maksyutenko2022}
Maksyutenko, P., Mart{\'\i}n-Dom{\'e}nech, R., Piacentino, E.~L., {\"O}berg,
  K.~I., \& Rajappan, M. 2022, ApJ, 940, 113

\bibitem[{Marcelino {et~al.}(2021)Marcelino, Tercero, Ag{\'u}ndez, \&
  Cernicharo}]{marcelino2021}
Marcelino, N., Tercero, B., Ag{\'u}ndez, M., \& Cernicharo, J. 2021, A\&A, 646,
  L9

\bibitem[{Marks {et~al.}(2023)Marks, Wang, Evseev, Kuznetsov, Antonov, \&
  Kaiser}]{marks2023a}
Marks, J.~H., Wang, J., Evseev, M.~M., {et~al.} 2023, ApJ, 942, 43

\bibitem[{McClure {et~al.}(2023)McClure, Rocha, Pontoppidan, Crouzet, Chu,
  Dartois, Lamberts, Noble, Pendleton, Perotti, {et~al.}}]{mcclure2023}
McClure, M.~K., Rocha, W., Pontoppidan, K., {et~al.} 2023, Nature Astronomy, 1

\bibitem[{McGuire(2022)}]{mcguire2022}
McGuire, B.~A. 2022, The Astrophysical Journal Supplement Series, 259, 30

\bibitem[{McGuire {et~al.}(2020)McGuire, Burkhardt, Loomis, Shingledecker, Lee,
  Charnley, Cordiner, Herbst, Kalenskii, Momjian, {et~al.}}]{mcguire2020}
McGuire, B.~A., Burkhardt, A.~M., Loomis, R.~A., {et~al.} 2020, ApJL, 900, L10

\bibitem[{McManus {et~al.}(2014)McManus, Martono, \& Vohs}]{McManus2014a}
McManus, J.~R., Martono, E., \& Vohs, J.~M. 2014, Catal. Today, 237, 157

\bibitem[{Minissale {et~al.}(2022)Minissale, Aikawa, Bergin, Bertin, Brown,
  Cazaux, Charnley, Coutens, Cuppen, Guzman, {et~al.}}]{minissale2022}
Minissale, M., Aikawa, Y., Bergin, E., {et~al.} 2022, ACS Earth Space Chem., 6,
  597

\bibitem[{Minissale \& Dulieu(2014)}]{Minissale2014}
Minissale, M. \& Dulieu, F. 2014, J. Chem. Phys., 141, 014304

\bibitem[{Muir {et~al.}(2020)Muir, Molina, Islam, Abdel-Rahman, \&
  Trenary}]{Muir2020a}
Muir, M., Molina, D.~L., Islam, A., Abdel-Rahman, M.~K., \& Trenary, M. 2020,
  PCCP, 22, 25011

\bibitem[{Mumma \& Charnley(2011)}]{mumma2011}
Mumma, M.~J. \& Charnley, S.~B. 2011, ARA\&A, 49, 471

\bibitem[{Nakamura {et~al.}(2022)Nakamura, Matsumoto, Amano, Enokido, Zolensky,
  Mikouchi, Genda, Tanaka, Zolotov, Kurosawa, {et~al.}}]{nakamura2022}
Nakamura, T., Matsumoto, M., Amano, K., {et~al.} 2022, Science, eabn8671

\bibitem[{Naraoka {et~al.}(2023)Naraoka, Takano, Dworkin, Oba, Hamase, Furusho,
  Ogawa, Hashiguchi, Fukushima, Aoki, {et~al.}}]{naraoka2023}
Naraoka, H., Takano, Y., Dworkin, J.~P., {et~al.} 2023, Science, 379, eabn9033

\bibitem[{Nazari {et~al.}(2022)Nazari, Meijerhof, van Gelder, Ahmadi, van
  Dishoeck, Tabone, Langeroodi, Ligterink, Jaspers, Beltr{\'a}n,
  {et~al.}}]{nazari2022b}
Nazari, P., Meijerhof, J., van Gelder, M., {et~al.} 2022, A\&A, 668, A109

\bibitem[{Nazari {et~al.}(2021)Nazari, van Gelder, Van~Dishoeck, Tabone,
  van’t Hoff, Ligterink, Beuther, Boogert, o~Garatti, Klaassen,
  {et~al.}}]{nazari2021}
Nazari, P., van Gelder, M., Van~Dishoeck, E., {et~al.} 2021, Astronomy \&
  Astrophysics, 650, A150

\bibitem[{Noble {et~al.}(2015)Noble, Diana, \& Dulieu}]{Noble2015}
Noble, J., Diana, S., \& Dulieu, F. 2015, MNRAS, 454, 2636

\bibitem[{Noble {et~al.}(2014)Noble, Theule, Duvernay, Danger, Chiavassa,
  Ghesquiere, Mineva, \& Talbi}]{Noble2014}
Noble, J., Theule, P., Duvernay, F., {et~al.} 2014, PCCP, 16, 23604

\bibitem[{Oba {et~al.}(2010)Oba, Watanabe, Kouchi, Hama, \&
  Pirronello}]{Oba2010a}
Oba, Y., Watanabe, N., Kouchi, A., Hama, T., \& Pirronello, V. 2010, ApJ, 722,
  1598

\bibitem[{{\"O}berg \& Bergin(2021)}]{oberg2021}
{\"O}berg, K.~I. \& Bergin, E.~A. 2021, Physics Reports, 893, 1

\bibitem[{{\"O}stblom {et~al.}(2005){\"O}stblom, Liedberg, Demers, \&
  Mirkin}]{Ostblom2005}
{\"O}stblom, M., Liedberg, B., Demers, L.~M., \& Mirkin, C.~A. 2005, J. Phys.
  Chem. B, 109, 15150

\bibitem[{Peck {et~al.}(1998)Peck, Mahon, \& Koel}]{Peck1998}
Peck, J.~W., Mahon, D.~I., \& Koel, B.~E. 1998, Surf. Sci., 410, 200

\bibitem[{Piacentino \& {\"O}berg(2022)}]{piacentino2022}
Piacentino, E.~L. \& {\"O}berg, K.~I. 2022, ApJ, 939, 93

\bibitem[{Pontoppidan {et~al.}(2003)Pontoppidan, Fraser, Dartois, Thi,
  Van~Dishoeck, Boogert, d'Hendecourt, Tielens, \& Bisschop}]{pontoppidan2003}
Pontoppidan, K., Fraser, H., Dartois, E., {et~al.} 2003, A\&A, 408, 981

\bibitem[{Potapov {et~al.}(2022)Potapov, Fulvio, Krasnokutski, J{\"a}ger, \&
  Henning}]{potapov2022a}
Potapov, A., Fulvio, D., Krasnokutski, S., J{\"a}ger, C., \& Henning, T. 2022, J.
  Phys. Chem. A, 126, 1627

\bibitem[{Potapov {et~al.}(2019)Potapov, Theul{\'e}, J{\"a}ger, \&
  Henning}]{Potapov2019b}
Potapov, A., Theul{\'e}, P., J{\"a}ger, C., \& Henning, T. 2019, ApJL, 878, L20

\bibitem[{Qasim {et~al.}(2019)Qasim, Fedoseev, Chuang, Taquet, Lamberts, He,
  Ioppolo, van Dishoeck, \& Linnartz}]{Qasim2019a}
Qasim, D., Fedoseev, G., Chuang, K.-J., {et~al.} 2019, A\&A, 627, A1

\bibitem[{Redhead(1962)}]{redhead1962}
Redhead, P. 1962, vacuum, 12, 203

\bibitem[{Remusat {et~al.}(2005)Remusat, Derenne, Robert, \&
  Knicker}]{remusat2005b}
Remusat, L., Derenne, S., Robert, F., \& Knicker, H. 2005, Geochimica et
  Cosmochimica Acta, 69, 3919

\bibitem[{Rubin {et~al.}(2019)Rubin, Altwegg, Balsiger, Berthelier, Combi,
  De~Keyser, Drozdovskaya, Fiethe, Fuselier, Gasc, {et~al.}}]{rubin2019}
Rubin, M., Altwegg, K., Balsiger, H., {et~al.} 2019, MNRAS, 489, 594

\bibitem[{Rze{\'z}nicka {et~al.}(2005)Rze{\'z}nicka, Lee, Maksymovych, \&
  Yates}]{Rzeznicka2005}
Rze{\'z}nicka, I.~I., Lee, J., Maksymovych, P., \& Yates, J.~T. 2005, J. Phys.
  Chem. B, 109, 15992

\bibitem[{Salter {et~al.}(2018)Salter, Stubbing, Brigham, \&
  Brown}]{Salter2018}
Salter, T.~L., Stubbing, J.~W., Brigham, L., \& Brown, W.~A. 2018, J. Chem.
  Phys., 149, 164705

\bibitem[{Salter {et~al.}(2019)Salter, Wootton, \& Brown}]{Salter2019}
Salter, T.~L., Wootton, L., \& Brown, W.~A. 2019, ACS Earth Space Chem., 3,
  1524

\bibitem[{Schaff \& Roberts(1994)}]{schaff1994}
Schaff, J.~E. \& Roberts, J.~T. 1994, J. Phys. Chem., 98, 6900

\bibitem[{Schaff \& Roberts(1998)}]{Schaff1998}
Schaff, J.~E. \& Roberts, J.~T. 1998, Langmuir, 14, 1478

\bibitem[{Schmitt-Kopplin {et~al.}(2010)Schmitt-Kopplin, Gabelica, Gougeon,
  Fekete, Kanawati, Harir, Gebefuegi, Eckel, \& Hertkorn}]{schmitt-kopplin2010}
Schmitt-Kopplin, P., Gabelica, Z., Gougeon, R.~D., {et~al.} 2010, PNAS, 107,
  2763

\bibitem[{Schneider {et~al.}(2019)Schneider, Caldwell-Overdier, Coppieters~‘t
  Wallant, Dau, Huang, Nwolah, Kasule, Buffo, Mullikin, Widdup,
  {et~al.}}]{Schneider2019}
Schneider, H., Caldwell-Overdier, A., Coppieters~‘t Wallant, S., {et~al.}
  2019, Monthly Notices of the Royal Astronomical Society: Letters, 485, L19

\bibitem[{Schwaner {et~al.}(1997)Schwaner, Fieberg, \& White}]{schwaner1997}
Schwaner, A., Fieberg, J.~E., \& White, J. 1997, J. Phys. Chem. B, 101, 11112

\bibitem[{Sephton(2002)}]{sephton2002}
Sephton, M.~A. 2002, Natural product reports, 19, 292

\bibitem[{Sexton {et~al.}(1985)Sexton, Hughes, \& Avery}]{sexton1985}
Sexton, B., Hughes, A., \& Avery, N. 1985, Surf. Sci., 155, 366

\bibitem[{Singh {et~al.}(2022)Singh, Zhu, La~Jeunesse, Fortenberry, \&
  Kaiser}]{Singh2022a}
Singh, S.~K., Zhu, C., La~Jeunesse, J., Fortenberry, R.~C., \& Kaiser, R.~I.
  2022, Nat. Commun., 13, 1

\bibitem[{Slayton {et~al.}(1995)Slayton, Aubuchon, Camis, Noble, \&
  Tro}]{Slayton1995}
Slayton, R., Aubuchon, C., Camis, T., Noble, A., \& Tro, N. 1995, J. Phys.
  Chem., 99, 2151

\bibitem[{Smith \& Kay(2018)}]{Smith2018}
Smith, R.~S. \& Kay, B.~D. 2018, J. Phys. Chem. B, 122, 587

\bibitem[{Smith {et~al.}(2016)Smith, May, \& Kay}]{smith2016}
Smith, R.~S., May, R.~A., \& Kay, B.~D. 2016, J. Phys. Chem. B, 120, 1979

\bibitem[{Solomun {et~al.}(1989)Solomun, Christmann, \&
  Baumg{\"a}rtel}]{Solomun1989}
Solomun, T., Christmann, K., \& Baumg{\"a}rtel, H. 1989, J. Phys. Chem., 93,
  7199

\bibitem[{Souda(2011)}]{Souda2011}
Souda, R. 2011, J. Chem. Phys., 135, 164703

\bibitem[{Sullivan {et~al.}(2016)Sullivan, Boamah, Shulenberger, Chapman,
  Atkinson, Boyer, \& Arumainayagam}]{sullivan2016}
Sullivan, K.~K., Boamah, M.~D., Shulenberger, K.~E., {et~al.} 2016, MNRAS, 460,
  664

\bibitem[{Tait {et~al.}(2005{\natexlab{a}})Tait, Dohn{\'a}lek, Campbell, \&
  Kay}]{tait2005b}
Tait, S.~L., Dohn{\'a}lek, Z., Campbell, C.~T., \& Kay, B.~D.
  2005{\natexlab{a}}, J. Chem. Phys., 122, 164707

\bibitem[{Tait {et~al.}(2005{\natexlab{b}})Tait, Dohn{\'a}lek, Campbell, \&
  Kay}]{tait2005a}
Tait, S.~L., Dohn{\'a}lek, Z., Campbell, C.~T., \& Kay, B.~D.
  2005{\natexlab{b}}, J. Chem. Phys., 122, 164708

\bibitem[{Tarczay {et~al.}(2017)Tarczay, F{\"o}rstel, G{\'o}bi, Maksyutenko, \&
  Kaiser}]{Tarczay2017}
Tarczay, G., F{\"o}rstel, M., G{\'o}bi, S., Maksyutenko, P., \& Kaiser, R.~I.
  2017, ChemPhysChem, 18, 882

\bibitem[{Theul{\'e} {et~al.}(2011{\natexlab{a}})Theul{\'e}, Borget, Mispelaer,
  Danger, Duvernay, Guillemin, \& Chiavassa}]{Theule2011a}
Theul{\'e}, P., Borget, F., Mispelaer, F., {et~al.} 2011{\natexlab{a}}, A\&A,
  534, A64

\bibitem[{Theul{\'e} {et~al.}(2011{\natexlab{b}})Theul{\'e}, Borget, Mispelaer,
  Danger, Duvernay, Guillemin, \& Chiavassa}]{Theule2011b}
Theul{\'e}, P., Borget, F., Mispelaer, F., {et~al.} 2011{\natexlab{b}}, A\&A,
  534, A64

\bibitem[{Thrower {et~al.}(2009)Thrower, Collings, Rutten, \&
  McCoustra}]{thrower2009a}
Thrower, J., Collings, M., Rutten, F., \& McCoustra, M. 2009, J. Chem. Phys.,
  131, 244711

\bibitem[{Tinacci {et~al.}(2022)Tinacci, Germain, Pantaleone, Ferrero,
  Ceccarelli, \& Ugliengo}]{tinacci2022}
Tinacci, L., Germain, A., Pantaleone, S., {et~al.} 2022, ACS Earth Space Chem.,
  6, 1514

\bibitem[{Toumi {et~al.}(2016)Toumi, Pi{\'e}tri, Chiavassa, \&
  Couturier-Tamburelli}]{toumi2016}
Toumi, A., Pi{\'e}tri, N., Chiavassa, T., \& Couturier-Tamburelli, I. 2016,
  Icarus, 270, 435

\bibitem[{Tsegaw {et~al.}(2017)Tsegaw, G{\'o}bi, F\"{o}rstel, Maksyutenko,
  Sander, \& Kaiser}]{Tsegaw2017}
Tsegaw, Y.~A., G{\'o}bi, S., F\"{o}rstel, M., {et~al.} 2017, J. Phys. Chem. A,
  121, 7477

\bibitem[{Turner {et~al.}(2019)Turner, Abplanalp, Bergantini, Frigge, Zhu, Sun,
  Hsiao, Chang, Meinert, \& Kaiser}]{Turner2019}
Turner, A.~M., Abplanalp, M.~J., Bergantini, A., {et~al.} 2019, Sci. Adv., 5,
  eaaw4307

\bibitem[{Turner {et~al.}(2016)Turner, Abplanalp, \& Kaiser}]{Turner2016}
Turner, A.~M., Abplanalp, M.~J., \& Kaiser, R.~I. 2016, ApJ, 819, 97

\bibitem[{Turner {et~al.}(2018)Turner, Bergantini, Abplanalp, Zhu, G{\'o}bi,
  Sun, Chao, Chang, Meinert, \& Kaiser}]{Turner2018}
Turner, A.~M., Bergantini, A., Abplanalp, M.~J., {et~al.} 2018, Nat. Commun.,
  9, 1

\bibitem[{Turner {et~al.}(2021)Turner, Chandra, Fortenberry, \&
  Kaiser}]{Turner2021a}
Turner, A.~M., Chandra, S., Fortenberry, R.~C., \& Kaiser, R.~I. 2021,
  ChemPhysChem, 22, 985

\bibitem[{Turner {et~al.}(2020)Turner, Koutsogiannis, Kleimeier, Bergantini,
  Zhu, Fortenberry, \& Kaiser}]{Turner2020}
Turner, A.~M., Koutsogiannis, A.~S., Kleimeier, N.~F., {et~al.} 2020, ApJ, 896,
  88

\bibitem[{Tylinski {et~al.}(2020)Tylinski, Smith, \& Kay}]{tylinski2020}
Tylinski, M., Smith, R.~S., \& Kay, B.~D. 2020, J. Phys. Chem. C, 124, 2521

\bibitem[{Tzvetkov {et~al.}(2004{\natexlab{a}})Tzvetkov, Koller, Zubavichus,
  Fuchs, Casu, Heske, Umbach, Grunze, Ramsey, \& Netzer}]{Tzvetkov2004b}
Tzvetkov, G., Koller, G., Zubavichus, Y., {et~al.} 2004{\natexlab{a}},
  Langmuir, 20, 10551

\bibitem[{Tzvetkov {et~al.}(2004{\natexlab{b}})Tzvetkov, Ramsey, \&
  Netzer}]{Tzvetkov2004a}
Tzvetkov, G., Ramsey, M., \& Netzer, F. 2004{\natexlab{b}}, Chem. Phys. Lett.,
  397, 392

\bibitem[{Ulbricht {et~al.}(2006)Ulbricht, Zacharia, Cindir, \&
  Hertel}]{ulbricht2006}
Ulbricht, H., Zacharia, R., Cindir, N., \& Hertel, T. 2006, Carbon, 44, 2931

\bibitem[{Van~Broekhuizen {et~al.}(2005)Van~Broekhuizen, Pontoppidan, Fraser,
  \& Van~Dishoeck}]{vanbroekhuizen2005}
Van~Broekhuizen, F., Pontoppidan, K., Fraser, H., \& Van~Dishoeck, E. 2005,
  A\&A, 441, 249

\bibitem[{Villadsen {et~al.}(2022)Villadsen, Ligterink, \&
  Andersen}]{villadsen2022}
Villadsen, T., Ligterink, N.~F., \& Andersen, M. 2022, A\&A, 666, A45

\bibitem[{Vinogradoff {et~al.}(2012)Vinogradoff, Duvernay, Farabet, Danger,
  Theul{\'e}, Borget, Guillemin, \& Chiavassa}]{vinogradoff2012b}
Vinogradoff, V., Duvernay, F., Farabet, M., {et~al.} 2012, J. Phys. Chem. A,
  116, 2225

\bibitem[{Wang {et~al.}(2022{\natexlab{a}})Wang, Kleimeier, Johnson, Gozem,
  Abplanalp, Turner, Marks, \& Kaiser}]{wang2022a}
Wang, J., Kleimeier, N.~F., Johnson, R.~N., {et~al.} 2022{\natexlab{a}}, PCCP,
  24, 17449

\bibitem[{Wang {et~al.}(2022{\natexlab{b}})Wang, Marks, Tuli, Mebel, Azyazov,
  \& Kaiser}]{Wang2022b}
Wang, J., Marks, J.~H., Tuli, L.~B., {et~al.} 2022{\natexlab{b}}, J. Phys.
  Chem. A

\bibitem[{Ward {et~al.}(2012)Ward, Hogg, \& Price}]{Ward2012}
Ward, M.~D., Hogg, I.~A., \& Price, S.~D. 2012, MNRAS, 425, 1264

\bibitem[{Wei {et~al.}(2004)Wei, Huang, \& White}]{Wei2004}
Wei, W., Huang, W., \& White, J. 2004, Surf. Sci., 572, 401

\bibitem[{Williams {et~al.}(2003)Williams, Bird, Sykes, Santra, \&
  Lambert}]{williams2003}
Williams, F.~J., Bird, D.~P., Sykes, E. C.~H., Santra, A.~K., \& Lambert, R.~M.
  2003, J. Phys. Chem. B, 107, 3824

\bibitem[{Xi {et~al.}(1994)Xi, Yang, Jo, Bent, \& Stevens}]{Xi1994}
Xi, M., Yang, M.~X., Jo, S.~K., Bent, B.~E., \& Stevens, P. 1994, J. Chem.
  Phys., 101, 9122

\bibitem[{Yang {et~al.}(2021)Yang, Sakai, Zhang, Murillo, Zhang, Higuchi, Zeng,
  L{\'o}pez-Sepulcre, Yamamoto, Lefloch, {et~al.}}]{yang2021}
Yang, Y.-L., Sakai, N., Zhang, Y., {et~al.} 2021, The Astrophysical Journal,
  910, 20

\bibitem[{Zahidi {et~al.}(1994)Zahidi, Castonguay, \& McBreen}]{zahidi1994}
Zahidi, E., Castonguay, M., \& McBreen, P. 1994, J. Am. Chem. Soc., 116, 5847

\bibitem[{Zhang {et~al.}(2023)Zhang, Yang, Zhang, Cox, Zeng, Murillo, Ohashi,
  \& Sakai}]{zhang2023}
Zhang, Z.~E., Yang, Y.-l., Zhang, Y., {et~al.} 2023, The Astrophysical Journal,
  946, 113

\bibitem[{Zheng \& Kaiser(2010)}]{Zheng2010}
Zheng, W. \& Kaiser, R.~I. 2010, J. Phys. Chem. A, 114, 5251

\bibitem[{Zhu {et~al.}(2019{\natexlab{a}})Zhu, Frigge, Bergantini, Fortenberry,
  \& Kaiser}]{Zhu2019a}
Zhu, C., Frigge, R., Bergantini, A., Fortenberry, R.~C., \& Kaiser, R.~I.
  2019{\natexlab{a}}, ApJ, 881, 156

\bibitem[{Zhu {et~al.}(2019{\natexlab{b}})Zhu, Frigge, Turner, Abplanalp, Sun,
  Chen, Chang, \& Kaiser}]{Zhu2019b}
Zhu, C., Frigge, R., Turner, A.~M., {et~al.} 2019{\natexlab{b}}, PCCP, 21, 1952

\bibitem[{Zhu {et~al.}(2018{\natexlab{a}})Zhu, Frigge, Turner, Kaiser, Sun,
  Chen, \& Chang}]{Zhu2018b}
Zhu, C., Frigge, R., Turner, A.~M., {et~al.} 2018{\natexlab{a}}, Chemical
  Commun., 54, 5716

\bibitem[{Zhu {et~al.}(2022{\natexlab{a}})Zhu, Kleimeier, Turner, Singh,
  Fortenberry, \& Kaiser}]{zhu2022a}
Zhu, C., Kleimeier, N.~F., Turner, A.~M., {et~al.} 2022{\natexlab{a}}, PNAS,
  119, e2111938119

\bibitem[{Zhu {et~al.}(2018{\natexlab{b}})Zhu, Turner, Abplanalp, \&
  Kaiser}]{zhu2018a}
Zhu, C., Turner, A.~M., Abplanalp, M.~J., \& Kaiser, R.~I. 2018{\natexlab{b}},
  ApJS, 234, 15

\bibitem[{Zhu {et~al.}(2022{\natexlab{b}})Zhu, Wang, Medvedkov, Marks, Xu,
  Yang, Yang, Pan, \& Kaiser}]{zhu2022c}
Zhu, C., Wang, H., Medvedkov, I., {et~al.} 2022{\natexlab{b}}, J. Phys. Chem.
  Lett., 13, 6875

\end{thebibliography}

\begin{appendix}

\section{Desorption parameters and molecular constants}

The analysis techniques used in this publication make use of a number of molecular constants, such as the moments of inertia, molecule mass, and symmetry factor, which are listed in Table \ref{tab:mol_const}. All the literature studies used in this work and the pre-exponential frequency factors and desorption energies determined with the Redhead-TST method are presented in Table \ref{tab:redhead-tst}. This table provides an overview of all relevant data to determine the desorption constants (that is, $T_{peak}$, $\beta$), but also information on the desorption surface and how the molecule was produced. 

\onecolumn 

\begin{longtable}{llrrrrr}
\caption{Molecular constants}
\label{tab:mol_const}\\
\hline\hline
	Molecule & Name & Mass & $I_{\rm x}$ & $I_{\rm y}$ & $I_{\rm z}$ & $\sigma$ \\
	& & amu & \multicolumn{3}{c}{amu $A^{2}$} & \\
\hline
\endfirsthead
\caption{continued.}\\
\hline\hline
	Molecule & Name & Mass & $I_{\rm x}$ & $I_{\rm y}$ & $I_{\rm z}$ & $\sigma$ \\
	& & amu & \multicolumn{3}{c}{amu $A^{2}$} & \\
\hline
\endhead
\hline
\endfoot
                                     C$_{2}$H$_{4}$ &                       Ethylene &  28 &   19.08 &   16.17 &    2.92 &  2 \\
                                         CH$_{2}$NH &                    Methanimine &  29 &   16.47 &   14.43 &    2.04 &  2 \\
                                      (CH$_{2}$-O)x &                            POM &  30 &   14.33 &   12.86 &    1.47 &  2 \\
                                     C$_{2}$H$_{6}$ &                         Ethane &  30 &   25.17 &   25.17 &    6.31 &  1 \\
                                     N$_{2}$H$_{2}$ &                        Diazine &  30 &   14.05 &   13.00 &    1.05 &  1 \\
                                   CH$_{3}$NH$_{2}$ &                    Methylamine &  31 &   22.75 &   21.84 &    4.96 &  1 \\
                                     N$_{2}$H$_{4}$ &                      Hydrazine &  32 &   20.55 &   18.71 &    3.55 &  1 \\
                                         NH$_{2}$OH &                  Hydroxylamine &  33 &   18.70 &   17.25 &    2.59 &  1 \\
\end{longtable}
\tablefoot{POM = Polyoxymethylene, AAT = Acetaldehyde Ammonia Trimer, AATF salt = Acetaldehyde Ammonia Trimer Formate salt. For the salts no moments of inertia or symmetry factors have been listed.}

\pagebreak
\scriptsize

\begin{landscape}
\begin{longtable}{llllllllllllll}
\caption{Literature and Redhead-TST desorption energies}
\label{tab:redhead-tst}\\
\hline\hline
	Molecule$^{a}$ & Surface$^{b}$ &  Precursor(s)$^{c}$ & Processing$^{d}$ & $T_{\rm peak}$  & $\beta$ & $\Lambda$ & $q_{rot,3D}$ & $q_{tr,3D}$ & $\nu_{lit}$ & $E_{\rm des,lit}$ & $\nu_{TST}$ & $E_{\rm des,Red}$ & Reference\\
	& & & & K & K s$^{-1}$ & pm & & & s$^{-1}$ & K & s$^{-1}$ & K & \\
\hline
\endfirsthead
\caption{continued.}\\
\hline\hline
	Molecule$^{a}$ & Surface$^{b}$ &  Precursor(s)$^{c}$ & Processing$^{d}$ & $T_{\rm peak}$  & $\beta$ & $\Lambda$ & $q_{rot,3D}$ & $q_{tr,3D}$ & $\nu_{lit}$ & $E_{\rm des,lit}$ & $\nu_{TST}$ & $E_{\rm des,Red}$ & Reference\\
	& & & & K & K s$^{-1}$ & pm & & & s$^{-1}$ & K & s$^{-1}$ & K & \\
\hline
\endhead
\hline
\endfoot
                                     C$_{2}$H$_{4}$ &               (P) ASW &                        C$_{2}$H$_{4}$ &                 No &  68 & 0.033 & 40 &      125 &    62 & 4.0E+15 &  2800 & 1.1E+16 &  2780 &         \citet{behmard2019} \\
                                     C$_{2}$H$_{4}$ &           Compact ASW &                        C$_{2}$H$_{4}$ &                 No &  60 & 0.033 & 43 &      103 &    55 & 4.0E+15 &  2400 & 7.1E+15 &  2420 &         \citet{behmard2019} \\
                                         CH$_{2}$NH &                    Ag &                      CH$_{3}$NH$_{2}$ &      5 keV e$^{-}$ & 115 & 0.008 & 30 &      201 &   109 &      -- &    -- & 5.3E+16 &  5110 &            \citet{Zhu2019b} \\
                                      (CH$_{2}$-O)x &                    Au &                   H$_{2}$CO:H$_{2}$CO &                 UV & 230 & 0.067 & 21 &      426 &   226 &      -- &    -- & 4.6E+17 & 10390 &        \citet{Butscher2016} \\
                                      (CH$_{2}$-O)x &                    Au &           H$_{2}$O:NH$_{3}$:H$_{2}$CO &            Thermal & 225 & 0.067 & 21 &      412 &   221 &      -- &    -- & 4.3E+17 & 10140 &        \citet{Duvernay2014} \\
                                     C$_{2}$H$_{6}$ &             MgF$_{2}$ &                              CH$_{4}$ &                 UV &  66 & 0.017 & 39 &      503 &    65 &      -- &  2060 & 4.5E+16 &  2840 &      \citet{Carrascosa2020} \\
                                     C$_{2}$H$_{6}$ &               (P) ASW &                        C$_{2}$H$_{6}$ &                 No &  67 & 0.033 & 39 &      514 &    66 & 6.0E+16 &  3000 & 4.7E+16 &  2840 &         \citet{behmard2019} \\
                                     C$_{2}$H$_{6}$ &           Compact ASW &                        C$_{2}$H$_{6}$ &                 No &  58 & 0.033 & 42 &      414 &    57 & 6.0E+16 &  2500 & 2.9E+16 &  2420 &         \citet{behmard2019} \\
\end{longtable}
\tablefoot{$^{a}$POM = Polyoxymethylene, AAT = Acetaldehyde Ammonia Trimer, AATF salt = Acetaldehyde Ammonia Trimer Formate salt. $^{b}$(P) = porous, (C) = crystalline $^{c}$P:C:E = CH$_{3}$CH$_{2}$CHO:CO:CH$_{3}$CH$_{3}$, P:A = CH$_{3}$COCOCH$_{3}$:CH$_{3}$CHO, G = HOCH$_{2}$CH(OH)CHO $^{d}$H = Hydrogenation, O = Oxygenation}
\end{landscape}

\normalsize

\twocolumn

\section{Elemental ratios}

\begin{figure*}[h!]
    \centering
    \includegraphics[width=0.95\textwidth]{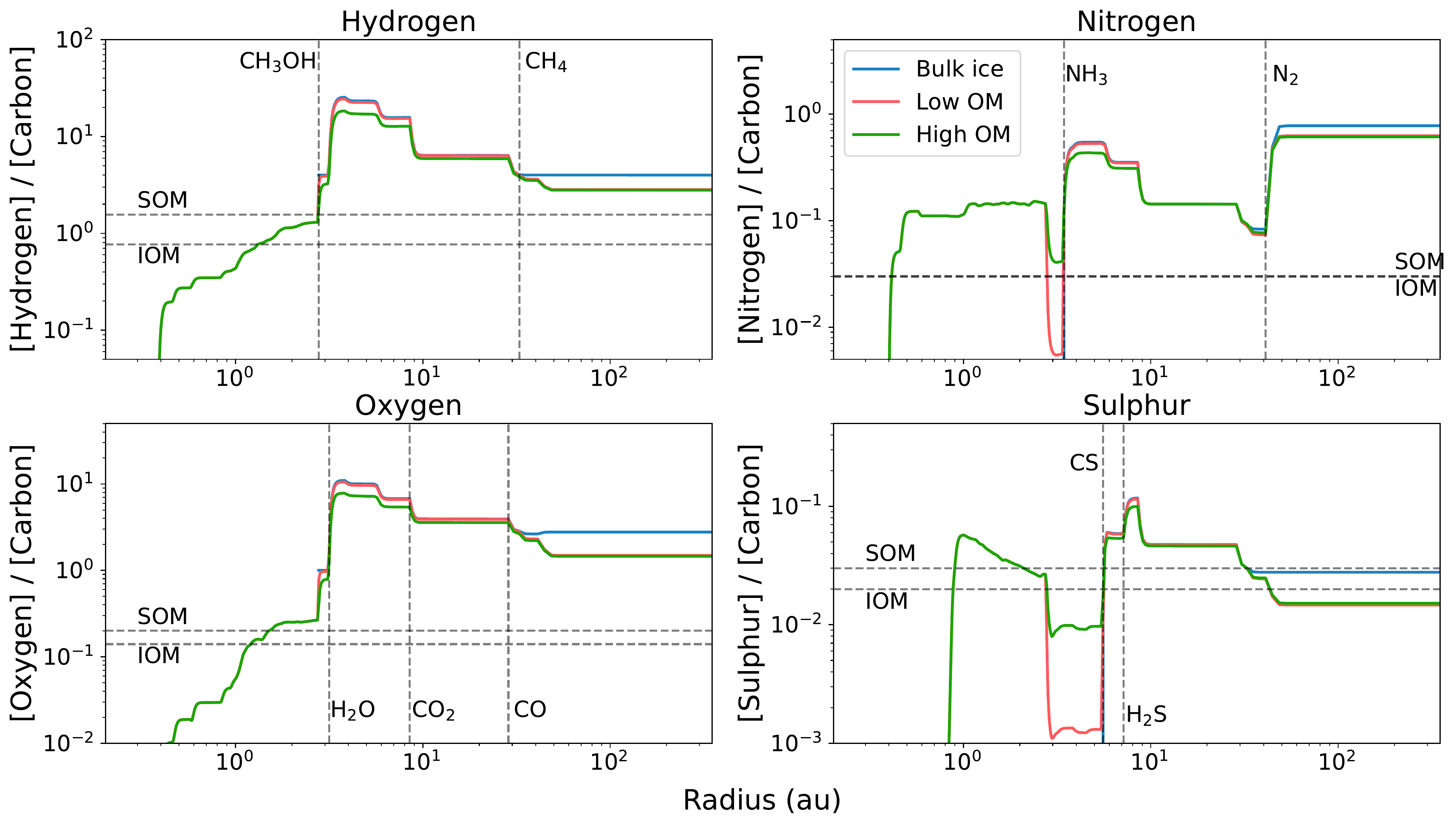}
    \caption{Ice elemental composition as [X]/[Carbon] for Hydrogen, Nitrogen, Oxygen, and Sulphur as a function of protoplanetary disk radius. Peak desorption radii for selected species are indicated in the plot (vertical dashed lines). Elemental compositions of Soluble Organic Matter (SOM) obtained from the Murchison meteorite \citep{schmitt-kopplin2010} and averaged Insoluble Organic Matter (IOM) \citep{alexander2017} are indicated (horizontal dashed lines). The desorption profiles are simulated with a first-order Polanyi-Wigner equation, variable surface coverages, and a heating rate of 1 K century$^{-1}$. An average protoplanetary disk temperature profile of $T(r) = 200 \times (r/ 1 AU)^{0.62}$ \citep{andrews2007} is used to determine the radii.}
    \label{fig:elem_rat_radius}
\end{figure*}

\begin{figure}[h!]
    \centering
    \includegraphics[width=0.45\textwidth]{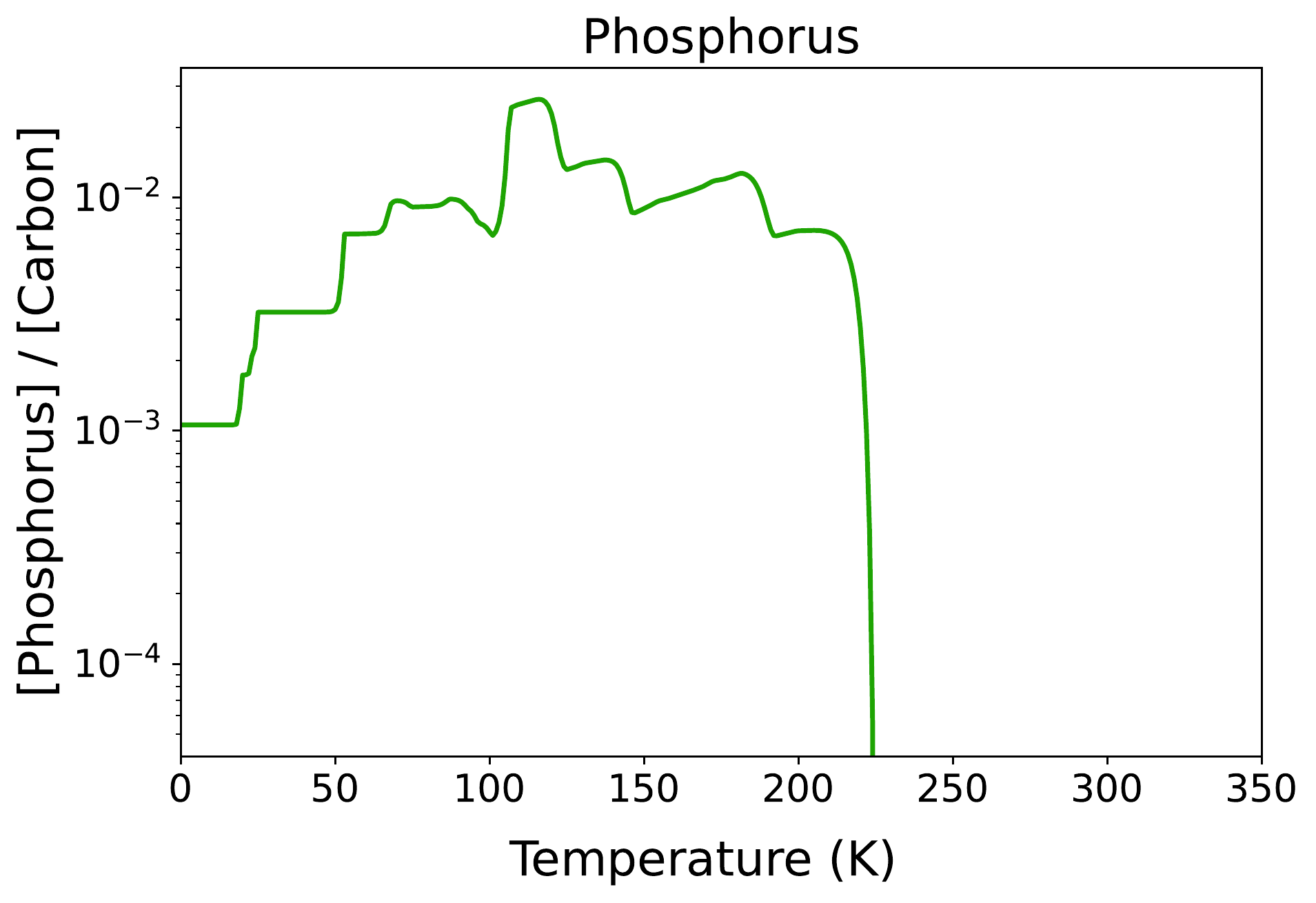}
    \caption{Ice elemental composition [Phosphorus]/[Carbon] as a function of temperature. The desorption profiles are simulated with a first-order Polanyi-Wigner equation, variable surface coverages, and a heating rate of 1 K century$^{-1}$.}
    \label{fig:phosphor}
\end{figure}

Based on the desorption parameters found in this work and assumed abundances, the ice elemental composition is determined throughout a protoplanetary disk, see Sect. \ref{sec:element}. The radial variations in elemental ratios are presented in Fig. \ref{fig:elem_rat_radius}. Figure \ref{fig:phosphor} presents the [Phosphorus]/[Carbon] ratio plotted against temperature.

\end{appendix}

\end{document}